\documentclass[%
  prb,%
  letterpaper,%
  twocolumn,%
  showpacs,%
  floatfix,%
  superscriptaddress%
]{revtex4-1}

\RequirePackage{graphicx,amsmath,amssymb,bm}
\RequirePackage{hyperref,verbatim}
\hypersetup{%
  breaklinks = {true},
  citecolor = {blue},
  colorlinks = {true},
  linkcolor = {red},
  pdfauthor = {\textcopyright\ Carlos Paez},
  pdfcreator = {\LaTeX\ and \flqq hyperref\frqq},
  pdffitwindow = {true},
  pdfmenubar = {true},
  pdfpagelayout = {SinglePage},
  pdfstartview = {Fit},
  pdftoolbar = {true},
  plainpages = {false},
}

\newcommand{\Fref}[1]{Fig.~\ref{#1}}

\renewcommand{\eqref}[1]{Eq.~(\ref{#1})}

\RequirePackage[usenames,dvipsnames]{xcolor}
\usepackage{soul}

\newcommand{\unicamplimeira}{Faculdade de Ci\^{e}ncias Aplicadas, Universidade
Estadual de Campinas, 13484-350 Limeira, SP
Brazil}

\newcommand{\minho}{Centro de F\'{\i}sica e Departamento de F\'{\i}sica, Universidade
do Minho, P-4710-057, Braga, Portugal}

\newcommand{\GRCNUS}{Centre for 2D Advanced Materials and Graphene Research Centre, Faculty of Science, National University
of Singapore, 6 Science Drive 2, Singapore 117546}

\begin{document}

\title{Electronic transport across linear defects in graphene}


\author{C. J. P\'aez} \email[Corresponding author:
]{carlos.gonzalez@fca.unicamp.br} \affiliation{\unicamplimeira}%
\author{J. N. B. Rodrigues} \affiliation{\GRCNUS}
\author{A. L. C. Pereira}%
\affiliation{\unicamplimeira}%
\author{N. M. R. Peres}%
\affiliation{\minho}%

\date{\today}

\begin{abstract}
  We investigate the low-energy electronic transport across grain boundaries in graphene ribbons 
  and infinite flakes. Using the recursive Green's function method, we calculate the electronic 
  transmission across different types of grain boundaries in graphene ribbons. We show results 
  for the charge density distribution and the current flow along the ribbon. We study linear defects 
  at various angles with the ribbon direction, as well as overlaps of two monolayer ribbon domains
  forming a bilayer region. For a class of extended defect lines with periodicity 3, an analytic 
  approach is developed to study transport in infinite flakes. This class of extended grain 
  boundaries is particularly interesting, since the $K$ and $K'$ Dirac points are superposed.
\end{abstract}

\maketitle

\section{Introduction}

The synthesis of graphene by chemical vapor deposition (CVD) on metal surfaces is the most widely used method for
producing graphene sheets.\cite{Li_Science:2009,Reina_NanoLett:2009,Kim_Nature:2009,Bae_NatureNanotech:2010} CVD
graphene, as any other solid grown by CVD, is especially prone to the formation of
 grain boundaries (GBs) and extended
defect lines, which hinder its electronic properties.\cite{Meyer_NL:2008,Lahiri_NatureNanotech:2010,HuaRV11,KwaZR11,Incze_APL:2011}

Graphene is being proposed for a variety of new
electronic devices. \cite{AvoD12,FioBI2014} However, the required high-quality
electrical properties are affected by the formation of
polycrystalline structures.\cite{TseBL12,YanXL14,TinsLR14,YazC14}  These
structures are practically unavoidable by the growth methods known so far.
\cite{HuaRV11,KwaZR11}  As such, the scattering problem of an
electron off a grain boundary (GB) becomes a theoretical and an experimental relevant one.
\cite{Gunlycke_PRL:2011,AyuJS14,RodPS13,RodPL12,GunW14}

The
$sp^{2}$ bonding structure of carbon atoms in graphene gives rise to extended topological defects that are typically
composed of pentagonal, heptagonal, and octagonal rings of carbon atoms, together with distorted
hexagons.\cite{Meyer_NL:2008,Lahiri_NatureNanotech:2010,HuaRV11,KwaZR11} GBs are in
general neither perfect straight lines nor periodic,
intercepting each other at random angles. However, periodic
straight GBs and defect lines can also be observed in graphene,\cite{Lahiri_NatureNanotech:2010} and
more interestingly they can be controllably synthesized at precise locations and orientations,\cite{Chen_PRB:2014,YanXL14}
lifting the prospects for the engineering of arrays of such defects that
would allow us to manipulate the electronic
valley degree-freedom in graphene.

GBs are known to strongly influence the properties of graphene,
namely its chemical, mechanical, and electronic
ones.\cite{Malola_PRB:2010,Liu_NanoLett:2010,Yazyev_PRB:2010,Yazyev_NatureMat:2010,Grantab_Science:2011}
GBs are expected to present different degrees of transparency to
electron transport, depending on their microscopic details and on the
relative orientations of the grains separated by
them.\cite{Yazyev_NatureMat:2010,Gunlycke_PRL:2011,Jiang_PLA:2011,Liwei_PRB:2012,RodPL12,RodPS13}
In fact, measurements of electronic mobilities of different CVD
samples, have shown that their electronic properties strongly depend
on the details of the CVD-growth
recipes.\cite{Reina_NanoLett:2009,Li_Science:2009,HuaRV11,Bae_NatureNanotech:2010}
Interestingly enough, and of direct relevance to our work,
 recent research has probed the electric properties
of single
GBs.\cite{Yu_NatureMat:2011,Jauregui_SSComm:2011,TseBL12}

In a revealing work, Yazyev {\it et al.}\cite{Yazyev_NatureMat:2010}
have studied electronic scattering from a wide variety of periodic
GBs. In that work, based on momentum conservation along
the periodic grain boundary, the authors have shed light on whether
 low-energy electrons travelling from one grain to the other may
feel a transport gap at the GB. Their conclusions were
also quantitatively corroborated by first-principles quantum transport
calculations (based on density functional theory and the non-equilibrium
Green's function formalism).

\begin{figure}[b]
  \begin{center}
    \includegraphics[width=0.95\columnwidth]{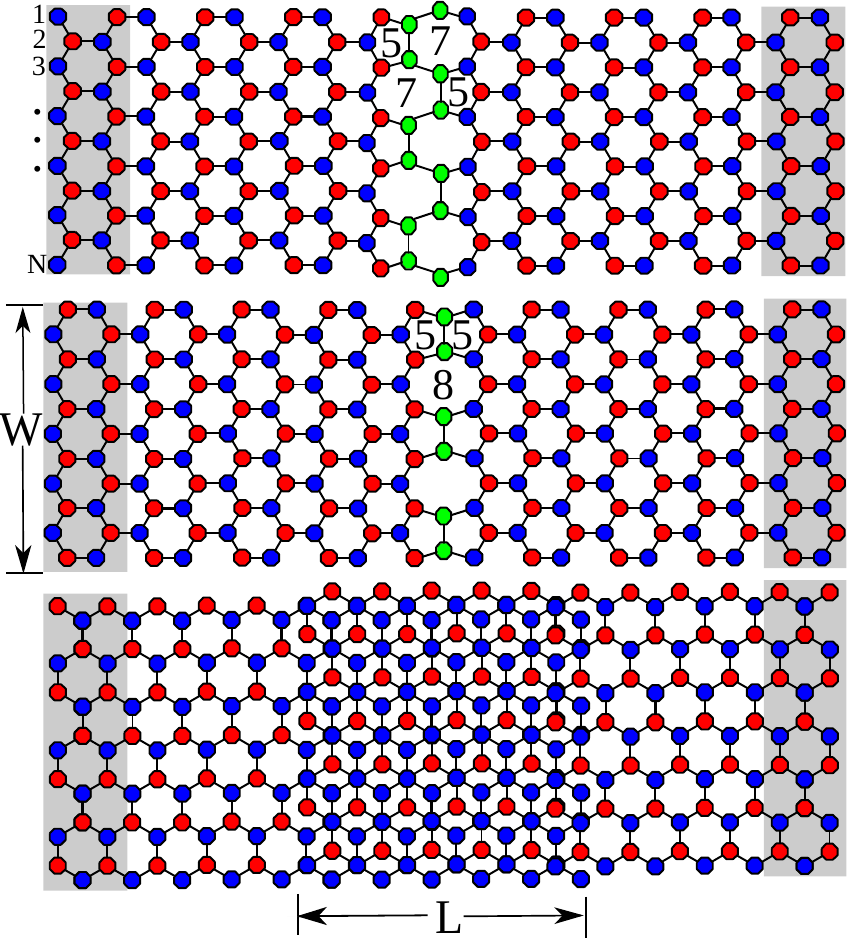}
    \caption{(color online) Schematic representation of the different
      linear defect structures considered as grain boundaries. On the
      top is the 5757 structure. In the middle we show the 585 linear
      defect. At bottom is the bilayer graphene of length L, formed by
      the overlap of two monolayer regions. The shadow areas represent
      the left and right semi-infinite contacts.}
    \label{fig:Fig1}
  \end{center}
\end{figure}

Our approach to the scattering problem due to GB's follows two different routes. 
In the first route we use the recursive Green's functions method to
numerically calculate the transmission through defect lines
in graphene ribbons. We map charge density over each sublattice site and also
the current density through the defects. Following the experiments,
\cite{TseBL12,YanXL14,TinsLR14}
we consider grain boundaries composed of
extended linear defects of type 585 (pentagons and octagons) and 5757
(pentagons and heptagons). Our results for the resistance across a linear
defect compare well to  recent experimental results.\cite{TseBL12} 
We also consider graphene ribbons with bilayer GBs: a spatial region where the grain
boundary is composed by the superposition of two monolayer domains, as shown in
\Fref{fig:Fig1}.  For this kind of
overlapping bilayer boundary, previous results have already shown
interesting conductance oscillations.\cite{GonSP10}  Here we show
that the transmission through these superpositions is reduced in
comparison to the transmission in the single-crystal domains. We also
present a spatial map of the current and the charge distribution
through these overlapped regions, which helps in the understanding of
the transport properties of  these systems.

In the second route, we will concentrate on a particular class of extended grain boundaries
briefly addressed by Yazyev {\it et al.}, namely, those grain boundaries with
periodicities such that both Dirac points (at each side of the grain
boundary) are mapped into the $\Gamma$ point of the projected Brillouin
zone. In such cases, there will be intervalley scattering of massless
Dirac fermions at the grain boundary.

We have chosen to investigate zigzag aligned extended linear defect lines separating two grains with the
same orientation (also referred to in the literature as degenerate, i. e., zero misorientation angle,
grain boundaries). Several such defect lines were proposed in the context of ab-initio works both
on graphene and on boron nitride: the t7t5 defect line\cite{Botello-Mendez_Nanoscale:2011} and the
7557 defect line\cite{Ansari_PCCP:2014} (see Fig. \ref{fig:Lit3PdfctLs}) are two such defect lines.
As we will see ahead, this defect line allows for valley scattering to occur and can thus be
regarded as a useful nanostructure for valleytronics circuits.

For this second type of problems, and in the context of graphene's single particle first neighbor
tight-binding model, we will use the transfer matrix formalism\cite{RodPS13} to analytically compute
the transmittance of electrons across these grain boundaries. The boundary condition seen by the
electrons at the grain boundary will be determined from its microscopic tight-binding model. In doing so,
we will follow the methodology developed for the cases of the pentagon-only, zz(558), and zz(5757)
defect lines.\cite{RodPL12,RodPS13} We will see that the obtained boundary condition explicitly introduces
intervalley scattering.

\section{Tight-binding model and transport formalism}
\label{sec:theory}

We consider graphene's tight-binding Hamiltonian
\begin{equation}
  \label{eq:hamiltonian}
  \begin{array}{c}
    H=-t_{i,j}\displaystyle\sum_{<i,j>}(c_{i}^{\dagger}c_{j}
    +\textsc{H.c.})
  \end{array}
\end{equation}
where $c_i$ ($c^{\dagger}_i$) annihilates (creates) an electron at site $i$ and $(i,j)$ stands for pairs of nearest-neighbor atoms.
We use the value of $t_{i,j}=2.7$ eV for the inplane nearest-neighbor hopping parameter and, when modeling the region of bilayer
graphene, we use $t_{ij,\perp}=0.381$ eV for the interlayer coupling.\cite{KuzCV09} The extended linear defects are constructed
by rearranging the positions of the atoms in the defect region. As shown in Fig. \ref{fig:Fig1}, this modifies the topology
of the lattice and thus changes pristine graphene's Hamiltonian in Eq. (\ref{eq:hamiltonian}).

Coherent transport across grain boundaries in graphene is studied
within the Landauer-B\"uttiker formalism, which relates the conductance
$G(E)$ at a given energy $E$ to the transmission function $T(E)$
between the contacts as
\begin{equation}
  \label{eq:conductance}
  \begin{array}{l}
    G(E) = G_0 T(E) \,,
  \end{array}
\end{equation}
with $G_0=2\frac{e^2}{h}\approx\frac{1}{12.5k\Omega}$. In the context of the previously referred first 
approach to scattering problems by GBs, the transmittance $T(E)$ is evaluated by means of the recursive Green's 
function approach using a two-terminal device configuration with contacts represented by the
semi-infinite ideal graphene leads
\begin{equation}
  \label{eq:transmission}
  \begin{array}{l}
    T=Tr\left[\Gamma_LG_S^{\dagger}\Gamma_{R}G_S\right ] \,,
  \end{array}
\end{equation}
where $G_S$ is the retarded Green's function of the system, given by\cite{Datta99}
\begin{equation}
  \label{eq:GR}
  \begin{array}{l}
    G_S=[E'I-H_S-\Sigma_{L}-\Sigma_{R}]^{-1} \,.
  \end{array}
\end{equation}
In these expressions $H_S$ is the Hamiltonian for the scattering region, $ \Sigma_{{L}({R})}=t^2g_{{L}({R})}$ stand for the
self-energies coupling the scattering region to the leads, while $E'$ is a shorthand for $E' = E+i\eta$, with $\eta \rightarrow 0$.
The self-energies
and the broadening function $\Gamma_{{L}({R})}=i\big(\Sigma_{{L}({R})}-\Sigma_{{L}({R})}^\dag\big)$\cite{Datta99} are calculated
from the electrode's Green's function $g_{L(R)}$ also obtained numerically using a recursive technique.\cite{LopLS85}

Charge and current are intimately related through the continuity equation. The connection with the Green's function arises 
from the quantum statistical average of the bond charge current operator, $\hat{J}_{ij}=\frac{e}{i\hbar}\big[t_{ji} 
c^{\dagger}_{j}c_i-t_{ij}c^{\dagger}_ic_{j}\big]$, which is related to the lesser Green's function
$G^<_{ji}(E)$.\cite{HauJ08,Datta99} In a steady state the bond charge current including spin degeneracy is:
\begin{equation}
  J_{ij}=I_0\int_{E_F^{-}}^{E_F^{+}} \textrm{d}E \Big[ t_{ji}G_{ij}^<(E)-t_{ij}G_{ji}^<(E) \Big] \,, \label{eq:current}
\end{equation}
where $E_{F}^{\pm} = E_F \pm eV/2$, while $I_0$ stands for the natural unit of bond charge current density being given by
$I_{0} = 2 e/h \approx 77.5 \, \mu$A/eV.

The lesser Green's function in the absence of interactions can be solved
exactly giving $G^<(E)=G_S(E)[\Gamma_Lf_L+\Gamma_Rf_R] G_S^\dagger(E)$, where $f_{L(R)}$ is the Fermi distribution of the left
(right) contact and $t_{ji}$ is the hopping parameter between sites $j$ and $i$. The bond current $ J_{ij}$ can
be visualized as a bundle of flow lines bunched together along a link joining the two sites.

Complementary to the current density, the charge density at site $j$ can also be expressed using the lesser Green's function as:
\begin{equation}
  \label{eq:cchargedensity}
  \rho_{(j)}=\frac{e}{2\pi
    i}\int_{E_F-eV/2}^{E_F + eV/2} dE{G_{j,j}^<(E)} \,.
\end{equation}

It is noteworthy that at low bias and low
temperature the charge density $\rho$, has the
same distribution of the local density of states (LDOS). Given that
we are interested in how charge and current
distributions are related, to keep explanations and figures as simple as
possible, we will refer from now on to LDOS as
charge distribution, with no loss of generality.

In addition to the Landauer-Buttiker formalism [see \eqref{eq:conductance}], it has been shown\cite{RodPS13,RodPL12}
how can we compute the low-energy limit of the conductance across this kind of
defect lines. Interestingly, at low temperatures, the conductance across a
defect line of size $W$ [see  \Fref{fig:Fig1}]
turned out to be linear in $K_F W$ and proportional to the transmittance
[see \eqref{eq:transmission}] close to Dirac point  ($E\rightarrow0$):
\begin{eqnarray}
  G(E)&=&W\frac{g_vg_s}{4\pi} \bigg\vert \frac{E}{\hbar
    v_F} \bigg\vert G_0T(E) \,. \label{eq:RodConduc}
\end{eqnarray}

The gate voltage $V_g$ is nothing more than the spatial potential distribution created by the substrate's charge distribution.
We have estimated $V_g$
for the GBs from the capacitor law
\begin{equation}
  V_g=\frac{q n d}{\epsilon A} \,,  \label{eq:gatevolt}
\end{equation}
where $n$ stands for the carrier density, $d$ is the thickness and $\epsilon$ is the dielectric constant of the substrate.
In order to convert the experimentally measured gate voltage into carrier density $n$ we use the relation $V_g=n/\alpha$,
where $\alpha=2.5 \times 10^{12}$ $m^{-2} V^{-1}$ is a geometry-related factor. From here onward, we will only consider the
carrier density to be $n=g_s g_v K_F^2 / 4\pi$, where $K_F$ is the momentum at the Fermi energy and $g_v$ ($g_s$) stands for
the valley (spin) degeneracy. Finally, as in graphene $E_F$ and $K_F$ are proportional at low-energy, $E_F=\hbar v_F K_F$, then
\begin{equation}
  \label{eq:gatevolt2}
  \begin{array}{l}
    V_g(E)=\frac{1}{\alpha}\frac{g_sg_v}{4\pi}(\frac{E
    }{\hbar v_F})^2\,,
  \end{array}
\end{equation}

From \eqref{eq:RodConduc} and \eqref{eq:gatevolt2} one can expect
that the resistance of a periodic defect line, at low temperatures and
in the linear regime, should behave as
\begin{eqnarray}
   R&=&\frac{1}{G} \propto\frac{1}{\sqrt{V_g}}\,.  \label{eq:resistance}
\end{eqnarray}
This square root dependence on $V_g$ should be clear from the experimental measurements of 
the resistivity across a grain boundary.

\section{Modified conductance quantization in the presence of the
  linear defect}

\begin{figure}[b]
  \begin{center}
    \includegraphics[width=1\columnwidth]{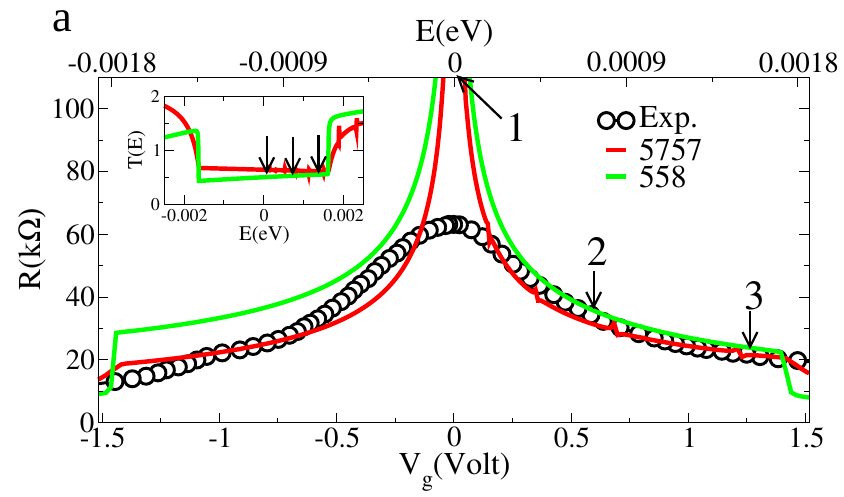}
    \includegraphics[width=1\columnwidth]{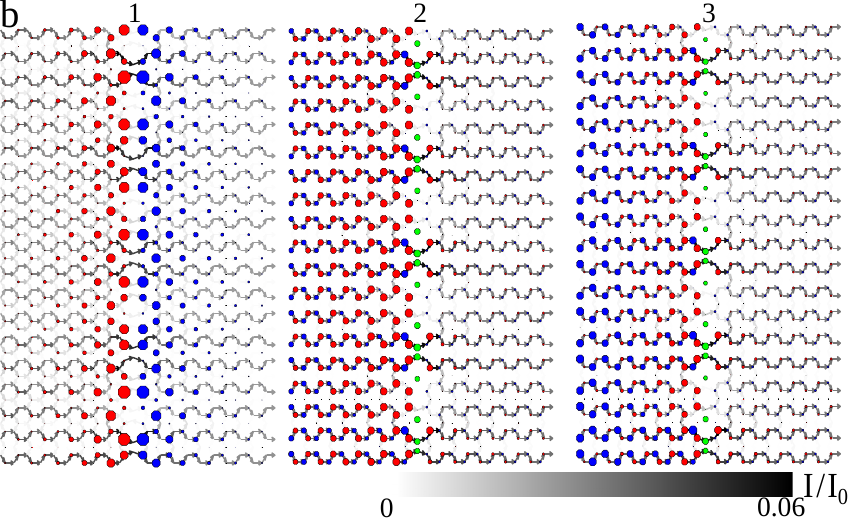}
    \caption{(color online)(a) Resistivity across the grain boundaries
      structures 5757 and 585, compared to the experimental results of
      Tsen \emph{et al.}\cite{TseBL12}.  Inset: Transmission
      function. Here we used an armchair ribbon with $g_s=2$,
      $g_v=1$, $N=56$ atoms along the width and making further
      correspondence to $W=1 \mu$m.  (b) Spatial distribution of
      charge densities and current densities over the GB for different
      energies: $E=10^{-7}$ eV, $E=7 \times 10^{-4}$ eV and
      $E=14 \times 10^{-4}$ eV . The charge densities are
      schematically represented here for a narrower ribbon. The
      current densities are evaluated at different sites.  The color
      of the arrow represents the magnitude of the electric current
      between any two neighboring sites, which are linearly normalized
      to the maximum value.}
    \label{fig:Fig2}
  \end{center}
\end{figure}

The study of scattering by extended defects is becoming increasingly more relevant, specially
after the recent work by Tsen {\it et
al.},\cite{TseBL12} where the authors made electric measurements across a single grain boundary. They have found that the
transport properties of these systems are strongly dependent on the GB's microscopic details.

Using \eqref{eq:RodConduc} and \eqref{eq:resistance} we calculate the resistance across two linear defects, the 5757 and the
585, and compare them with the experimental result from Tsen \emph{et al.} (black circles) -- \Fref{fig:Fig2}(a). We verify
that they agree to a good extent for $V_g$ not too close to the Dirac point. The disagreement (at low carrier densities)
between the experiment and our prediction is  due to the effect of puddles,
which are not taken into account in our
calculations, and dominate the bulk region of the device. Note that the effect of puddles is increasingly relevant when one 
approaches energies very close to the Dirac
point. In order to obtain the above results, we have considered an armchair ribbon with $N=8132$ sites ($g_s=2$, $g_v=1$
and $W\approx1 \mu$m). The electronic properties at low-energy regimes, are obtained by a rescaling of the electronic properties
of an armchair ribbon with $N=500$ sites.

The transport properties of these different linear defects are qualitatively similar. In the inset of \Fref{fig:Fig2}(a) we
show the electronic transmission and how it depends on the detailed geometry of the GB. In the continuum low-energy limit,
both the 585 and the 5757 defects have a metallic behavior with a flat band crossing the Fermi level.

The spatial distribution of charge density (for different energies) of a 585 linear defect is shown in \Fref{fig:Fig2}(b). The
density on each atomic site is represented as a disk. The different colors (red and blue) identify the sublattice, while the
magnitude of the disk's radius is proportional to the charge density at that site. We can see that the closer we are to the
Dirac point ($E=10^{-7}$eV) the more localized the charge is in the region of the GB. We also plot on \Fref{fig:Fig2}(b) the
distribution of charge density for higher energies. In this case, a higher dispersion of the charge is apparent mainly before
the 585 structure: the line defect acts as a potential wall.

The corresponding current densities are also shown in \Fref{fig:Fig2}(b), being evaluated at different sites using
\eqref{eq:current}. The color of the arrow represents the magnitude of the electric current between any two neighboring
sites, which are linearly normalized to the maximum value, according to the greyscale bar. For all the plotted energies,
we observe that before and after the linear defect, the current flows in a specific horizontal pattern along armchair
paths (streamlines) skipping some horizontal bonds, in accordance with recent ab-initio calculations of the current
densities in pristine armchair graphene ribbons.\cite{WilWE14} Here, with the linear defect, it is interesting to
observe how the current gives priority to some paths within the defect line, also in a periodical pattern. One can also
see that the current density is smaller for the first represented energy (i. e., $E=10^{-7}$ eV), as would be expected
due to the localized nature of the charge density around the defect for this energy.

\section{Linear defect orientation}

\begin{figure}[h]
  \begin{center}
    \includegraphics[width=1\columnwidth]{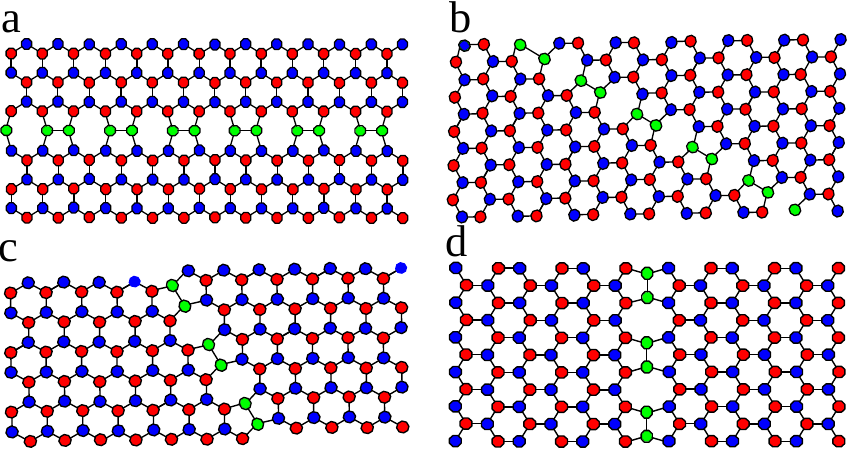}
    \includegraphics[width=1\columnwidth]{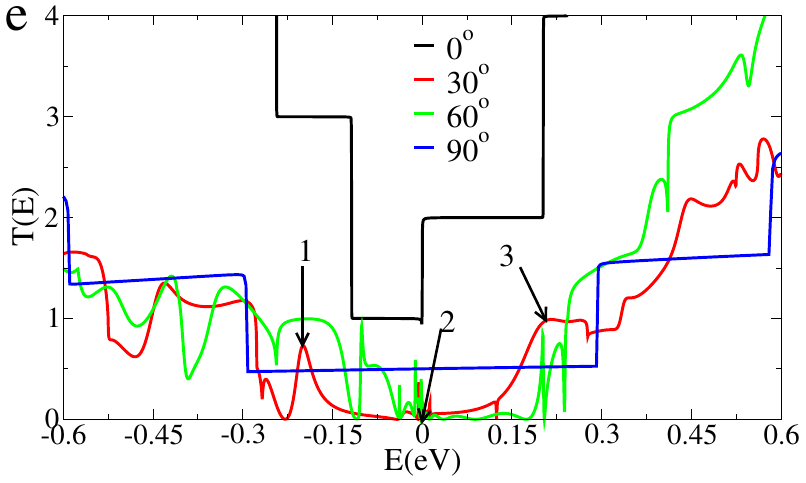}
    \caption{(color online) Schematic representation of a linear
      defect 585 at different orientation angles (a) $0^{\circ}$, (b)
      $30^{\circ}$, (c) $60^{\circ}$ and (d) $90^{\circ}$. (e) Transmission probability
      for each angle.  The ribbon has $N = 56$ sites in the width.}
    \label{fig:Fig3}
  \end{center}
\end{figure}

The transmission across linear defects is known to significantly depend on their orientation angle.\cite{VancL14,VancL13,RodPL12}
In \Fref{fig:Fig3}(a)-(d), we show schematic representations of four different orientations of a 585 extended defect in a graphene
ribbon: $\theta = 0^{\circ}, 30^{\circ},60^{\circ}$ and $90^{\circ}$. In \Fref{fig:Fig3}(e) we plot the transmission probability as a 
function of the energy for these different angles of incidence at the defect. For all cases, we see that the particle-hole symmetry 
is broken due to the translation symmetry breaking introduced by the defect.

\begin{figure}[h]
  \begin{center}
    \includegraphics[width=1\columnwidth]{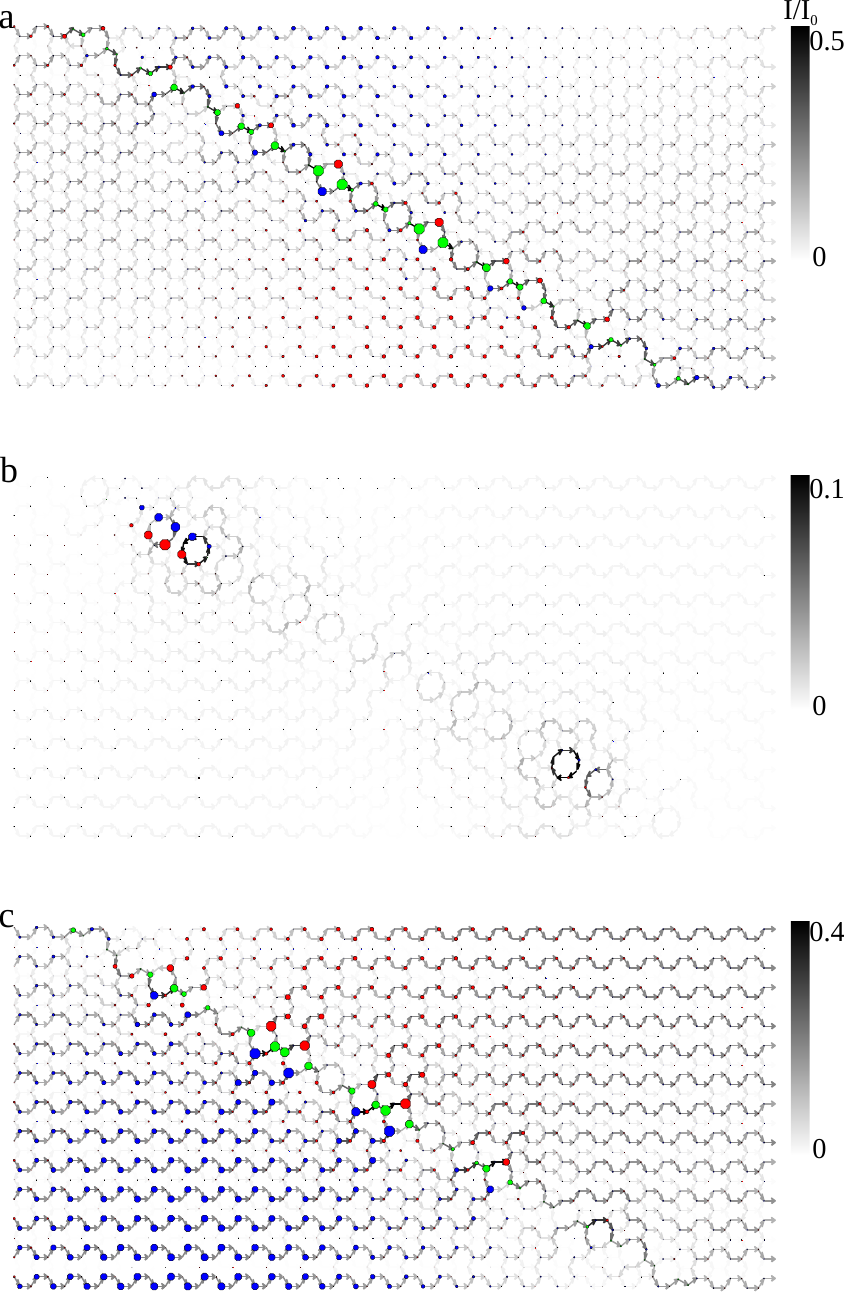}
    \caption{(color online) Spatial distribution of charge densities
      and current densities over the linear defect 585 with a
      orientation angle of $\theta=30^{\circ}$, to corresponding to the
      energies indicated by the numbered arrows in \Fref{fig:Fig3}(b):
      $E=-0.21$ eV, $E=1 \times 10^{-7}$ eV and $E=0.21$ eV.}
    \label{fig:Fig4}
  \end{center}
\end{figure}

For $\theta=0^{\circ}$, the presence of the defect line located in the
middle of the ribbon does not alter the metallic character
observed in the transmission spectrum of a pristine armchair
ribbon of the same width.\cite{BahPS11}  \Fref{fig:Fig3}(e) also
shows that for $\theta=30^{\circ}$ and $60^{\circ}$ there are regions of vanishing
transmission (opening of transport gap) close to the Dirac point. To
further investigate the origins of these oscillations in the
transmission, in \Fref{fig:Fig4} we map the charge and current density
distributions for the selected energies  indicated (by the arrows 1, 2 and 3) in
\Fref{fig:Fig3}(e).

The first energy (arrow 1) corresponds to a transmission resonance at
an energy $E=-0.21$ eV, which is typical of resonant tunneling
structures, where the continuum background is strongly suppressed at
the discrete state localization.  The associated charge and current
density distribution are shown in \Fref{fig:Fig4}(a). A high current
density exactly following the defect line reveals its metallic
character.  Close to the contacts and far from the defect line, the
current flow splits again into streamlines.\cite{WilWE14} A similar
behavior (high current density located along the linear defect) is
found for resonances at angle of $60^{\circ}$ (not shown here). We also
found that the resonances of the first channel in the angles $30^{\circ}$
and $60^{\circ}$ are robust structures independent from W.  The present
results suggest an image of Fano-type resonances, a finite coupling
between the localized state associated to the linear defect and the
delocalized continuum states associated with the armchair ribbon.

In \Fref{fig:Fig4}(b), we show the charge and current density
distribution for the second selected energy [arrow 2 in \Fref{fig:Fig3}(e)], corresponding
to a vanishing transmission close to the Dirac point. Interference in
charge and current density on both sides of the linear defect can be
seen, as well as on the linear defect.  For this low energy we
observe that the charge distribution is highly localized in part of
the defect. The current seems not to flow, with its maximum local
values circulating around the octagons where the charge is
concentrated and around octagons symmetrically
positioned with respect to the middle of the linear defect. Note
that the directions of the small arrows representing the local current
flow around the octagons is different for \Fref{fig:Fig4}(b) (local loopings)
and \Fref{fig:Fig4}(a) (net flow).  The backscattering is evident.
Therefore we conclude that at low-energies, i. e., close to Dirac point, the
suppression of the transmission, at the first electron-like plateau, is
due to charge localization and backscattering of a defect-related mode
of the 585 defect line.

\Fref{fig:Fig4}(c) corresponds to an energy value of $E = 0.21$ eV. One
can see that the current throughout the ribbon is not uniform and
forms ambiguous paths. At this energy, the current-density amplitude
also flows across the defect, being greater in the edge region than in
the center of the ribbon. The flow is mostly perpendicular
to the linear defect. Note that an electron can travel between the
source and the drain via many different transport channels. The local
electric current profile at a given energy is nothing more than the 
result from the interference between all the active transport channels 
at that energy. In particular, the existence of current loops for some defect
orientations (see panels of \Fref{fig:Fig4}), simply results from particular
interference patterns arising from the different blockade of distinct transport 
channels by the linear defect. Moreover, the particular local current patterns that are observed
result from the interplay between the different components of the nanostructure: linear defect
topology and orientation, edges type and width of the graphene ribbon. 
In \Fref{fig:Fig3}(b), for $\theta=90^{\circ}$, the first
plateau does not present interference oscillations, the transmission is
reduced in the vicinity of the Dirac point, due to the coupling
of extended states at the edges with localized ones at the defect line.
The oscillations at high energy range are simply Fabry-Perot
interference effects.

\section{Bilayer graphene as a grain boundary}

\begin{figure}[h]
  \begin{center}
    \includegraphics[width=0.9\columnwidth]{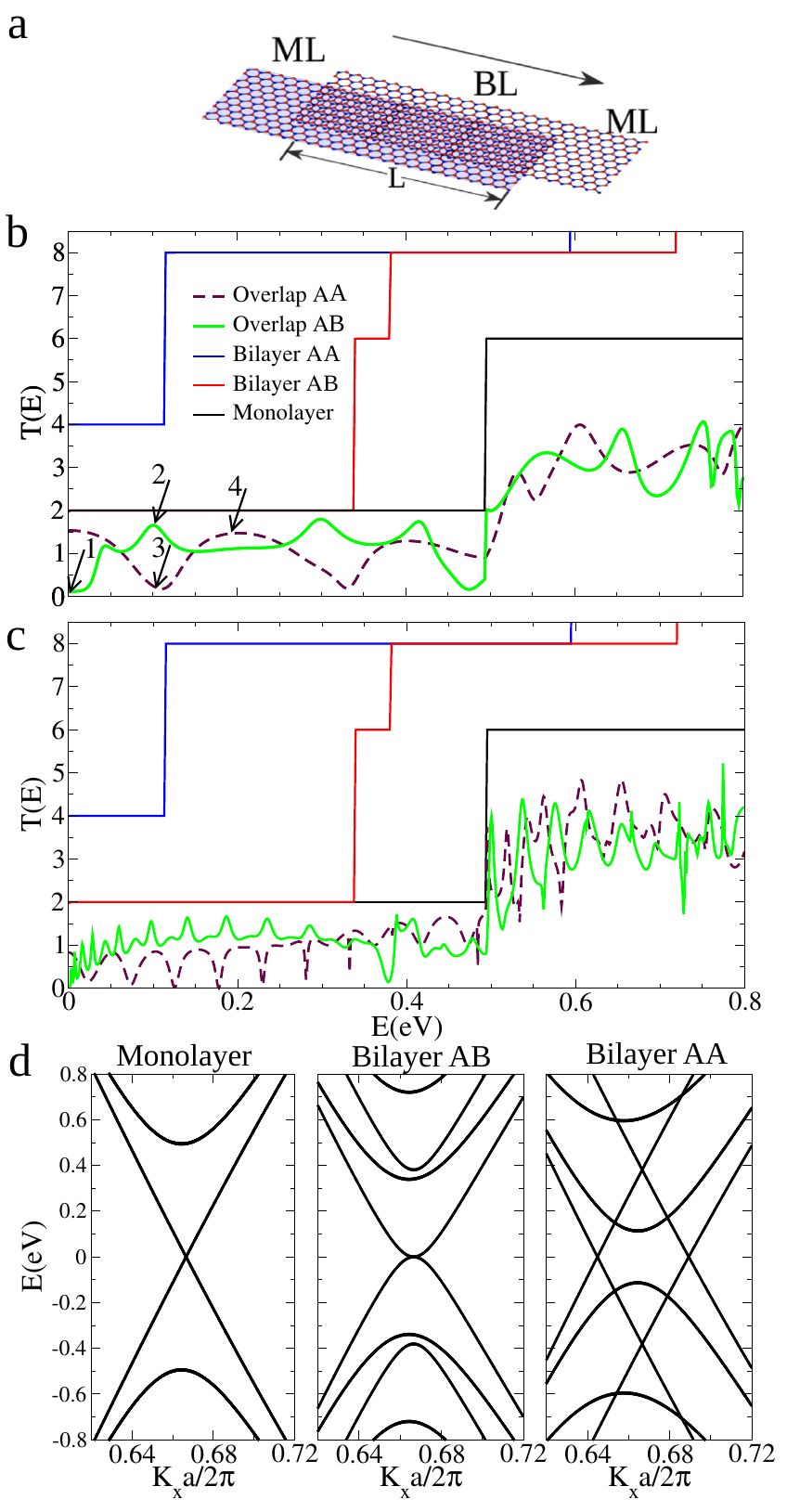}
    \caption{(color online) (a) Schematic representation of the overlap between
    two graphene monolayers, forming a bilayer region of length L.
    (b) Transmission throughout a bilayer region of length L corresponding to 80
    atoms superposed. Both AA and AB stackings are considered for the overlap
    (bilayer region). For comparison, it is shown the transmission through a
    pristine monolayer and pristine bilayers AA and AB of same width.
    (c) Same as in (b), showing now the transmission throughout a bilayer 
    region of length L corresponding to 320 atoms superposed.
    (d) Band structures for pristine monolayer and pristine bilayers AA and AB of same width.}
    \label{fig:Fig5}
  \end{center}
\end{figure}

In this section we investigate the electronic transport properties of a grain
boundary defined by an overlap between two semi-infinite monolayer graphene
regions, forming a bilayer region as represented in \Fref{fig:Fig5}(a).
Such overlaps have been experimentally observed.\cite{TseBL12}  Here, to focus
on the effects of the bilayer region on the transmission, we consider
periodical boundary conditions, avoiding edge localization effects.

Figure \ref{fig:Fig5}(b) and (c) shows the transmission T(E) as a function of
energy across a bilayer region of length $L$ corresponding to 80 and 320 atoms
superposed, respectively. We consider both the AB (Bernal) and AA stacking
cases for the overlap regions. For both of them oscillations in the transmission are observed,
with their frequency increasing for increasing overlap length $L$, in agreement
with previous calculations for similar overlaps.\cite{GonSP10} This 
can be qualitatively understood by remembering that in such systems, the transmittance 
is set by the wave-function matching at the monolayer-bilayer interfaces. A monolayer 
eigenstate incoming from the left is going to be partially transmitted into the bilayer
region and partially reflected back into the left monolayer. The portion of the 
wave-function transmitted into the bilayer region is going to propagate (acquiring a complex 
phase) until the second interface (bilayer-monolayer) and there it will undergo a similar 
scattering process: it is going to be partially reflected back to the bilayer and partially 
transmitted into the right monolayer. The resultant standing wave, in particular, the weight 
associated with each of its components (channels), is going to be the direct result of this 
interference process and will thus strongly depend on the length of the bilayer region and on 
the wave-number associated with each of those channels. The phases acquired by each of the 
wave-function's components of the bilayer region at the second interface are going to be 
smaller for shorter bilayer regions (i. e., shorter $L$). In such cases, oscillations in the 
transmittance will require a greater change of the eigenstates' wave-number, i. e., 
a greater increase in energy, as observed in both panels of \Fref{fig:Fig5}.
For comparison, we also show in \Fref{fig:Fig5}(b) and (c) the transmission
through a pristine monolayer and pristine bilayers AA and AB (considering the same
width, 40 atoms, to which periodic boundary conditions are applied). One can see
that the transmission through the pristine monolayer and bilayers is always higher
than the transmission throughout the overlapped region. This is due to the presence of 
the interfaces, that act as scattering centers decreasing the system's transmission.
Figure \ref{fig:Fig5}(d) shows the band structure for pristine monolayer and pristine
bilayers AA and AB (of same width and also with periodical boundary conditions),
which helps in understanding the origin of the plateaus in the transmission for each case.
At low energies, in the case of AB stacking there is only one conducting channel, whereas in AA stacking 
there are always two conducting channels for each valley. This partially explains why in general
the low-energy transmission for the AB bilayer structure is smaller than that for the AA bilayer
structure. But in addition to this, we can easily check that the boundary condition at a monolayer-bilayer 
AA interface can be completely satisfied at low energies without the need for reflected components
in the monolayer region. The same does not happen for the case of the monolayer-bilayer AB interface. 
Therefore, the upper bound for the transmission at low energies is smaller in the bilayer AB case than
in the bilayer AA case. Nevertheless, and by appropriately choosing the bilayer region length, $L$, we
can still make bilayer AB case's low-energy transmission higher than that of the bilayer AA case.   

\begin{figure*}[htp!]
  \begin{center}
    \includegraphics[width=1.85\columnwidth]{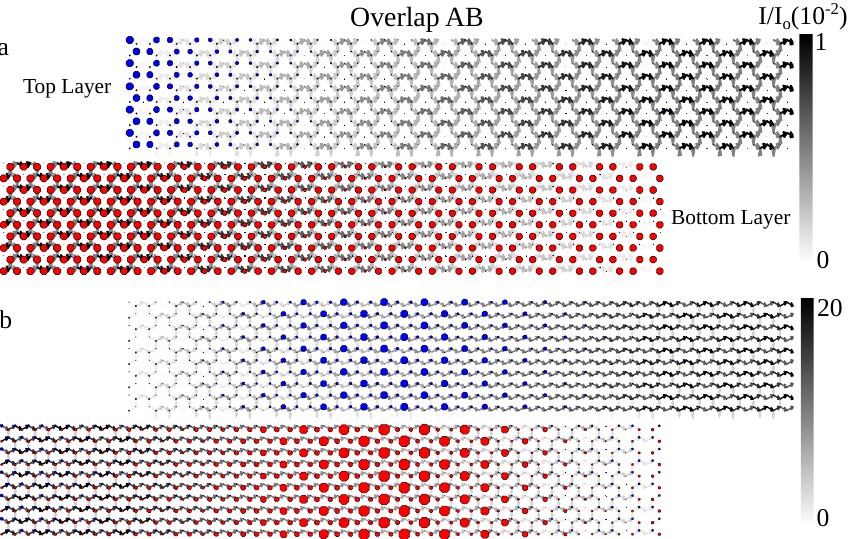}
    \caption{(color online) Charge and current density distributions
      for the nanostructure composed of two graphene (monolayer)
      ribbons that partially overlapped (AB-stacked in the
      overlap/bilayer region). The two panels stand for two
      different energies: (a) $E=0.001$ eV, which corresponds to a
      minimum of transmission (indicated by the arrow 1 in
      \Fref{fig:Fig5}). (b) $E=0.1$ eV, which corresponds to a maximum
      of transmission (indicated by the arrow 2 in \Fref{fig:Fig5}).}
    \label{fig:Fig6}
  \end{center}
\end{figure*}

In \Fref{fig:Fig6} we map the spatial distribution of charge density and local
current density on each of the overlapping (AB stacking) graphene monolayer ribbons.
Figure \ref{fig:Fig6} (a) and (b) show such maps corresponding to the energies $E=0.001$ eV
and $E=0.1$ eV, indicated by arrows 1 and 2 in \Fref{fig:Fig5}, respectively
a minimum and a maximum values of transmission in the low-energy limit.
For $E=0.001$ eV, i. e. in \Fref{fig:Fig6}(a), the charge density is localized mainly on the
left semi-infinite monolayer, which corresponds to the bottom layer in the bilayer region. On
right semi-infinite monolayer (i. e., the top layer of the bilayer region), the charge density 
rapidly decreases from the left to the right. In particular, its charge density is 
insignificant in the monolayer region. In the regions with non-vanishing charge density of each layer, 
the charge density typically concentrates on only one sublattice: the non-dimer sublattice. 
This effect comes from the sublattice asymmetry introduced by the AB-stacking in bilayer graphene.\cite{CasNM10,PaeBP2014}
Moreover, for this energy, the current density is very low on both layers. On the other hand, for $E=0.1$ eV
(which corresponds to a peak in the transmission), we can see in \Fref{fig:Fig6}(b) that again the
charge seems to be polarized on the non-dimer sublattice in the central part of the bilayer region.
However, a careful observation of other parts of the bilayer region shows that the charge is more 
homogeneously distributed over both sublattices there, thus allowing electron hopping between sites and 
between layers,\cite{McC06,KosA06} as observed in the pattern of the zigzag current density streamlines.

\begin{figure*}[t!]
  \begin{center}
    \includegraphics[width=1.85\columnwidth]{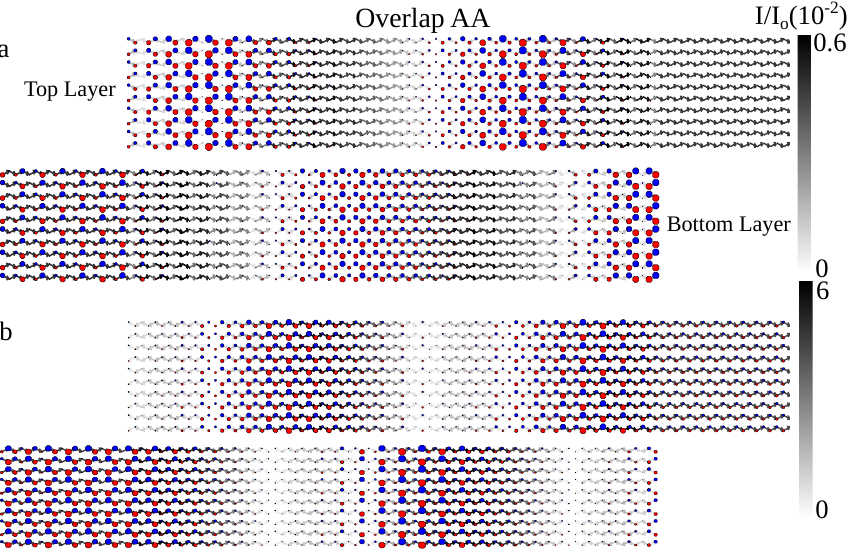}
    \caption{(color online) Charge and current density distributions
      for the nanostructure composed of two graphene (monolayer)
      ribbons that partially overlapped (AA-stacked in the
      overlap/bilayer region). The two panels stand for two different
      energies: (a) $E=0.11$ eV, which corresponds to a low
      transmission (indicated by the arrow 3 in \Fref{fig:Fig5}). (b)
      $E=0.2$ eV, which corresponds to a maximum of transmission
      (indicated by the arrow 4 in \Fref{fig:Fig5}).}
    \label{fig:Fig7}
  \end{center}
\end{figure*}

Similarly, in \Fref{fig:Fig7} we show the charge and current density distribution
for the case of the two monolayers with an AA-stacking overlap region. 
Figure \ref{fig:Fig7}(a) corresponds to $E=0.11$ eV (the minimum in transmission
indicated by the arrow 3 in \Fref{fig:Fig5}), while \Fref{fig:Fig7}(b) corresponds to $E=0.2$ eV
(the resonance in transmission indicated by the arrow 4 in \Fref{fig:Fig5}).
For both energies, there is a clear charge wave along the bilayer length with charge
oscillating between the two layers. Similarly, current also oscillates between layers. However comparing
charge and current densities in each layer, one can see an interesting behavior:
for the energy corresponding to low transmission, \Fref{fig:Fig7}(a) shows that there is
a clear imbalance, since the electronic charge density and the current
density are concentrated on different parts of the bilayer region. On the other hand, in
\Fref{fig:Fig7}(b), i. e., for the energy corresponding to high transmission, we again observe a
charge and current oscillation between layers along the length of the bilayer region, but now these are 
in phase, with the maximum current density spatially coinciding with the maximum charge density. Also 
note the higher current densities associated with the later energy (see different current grey scale bar).

\section{Transport across three periodic grain boundaries: transfer matrix approach}

\label{sec:TransfMatrixFormalism}

In this section we use the transfer matrix formalism\cite{RodPS13} to study the electronic transport across extended grain
boundaries in the scope of the single particle first neighbor tight-binding model of (infinite) monolayer graphene. This method
is entirely equivalent to the recursive Green's function numerical method used up to this point of the manuscript.
It reduces the electronic scattering problem to a set of matrix manipulations easy to work out by any computational
algebraic calculator, and thus give rise to an analytic solution of the problem.

We will concentrate on a particular class of zigzag aligned periodic GBs that separate grains with the same orientation
(also known as zero misorientation angle GBs) and have a periodicity that allows for intervalley scattering of
low-energy electrons. When the periodicity of the zigzag aligned GBs is a multiple of 3, both Dirac points (as well as
the $\Gamma$-point) are mapped into $k_{x} a = 0$ -- see panels (c) and (d) of \Fref{fig:FBZs}. Therefore, and in
contrast with what happens for the pentagon-only, 585 and 5757
GBs,\cite{Gunlycke_PRL:2011,Jiang_PLA:2011,Liwei_PRB:2012,RodPL12,RodPS13} linear momentum conservation does not
forbid low-energy electrons from scattering between valleys. Nevertheless, and if we want to know how much
intervalley scattering is a particular GB going to generate, we need to explicitly compute the boundary condition
matrix originating from its tight-binding microscopic model.

\begin{figure}
  \includegraphics[width=0.98\columnwidth]{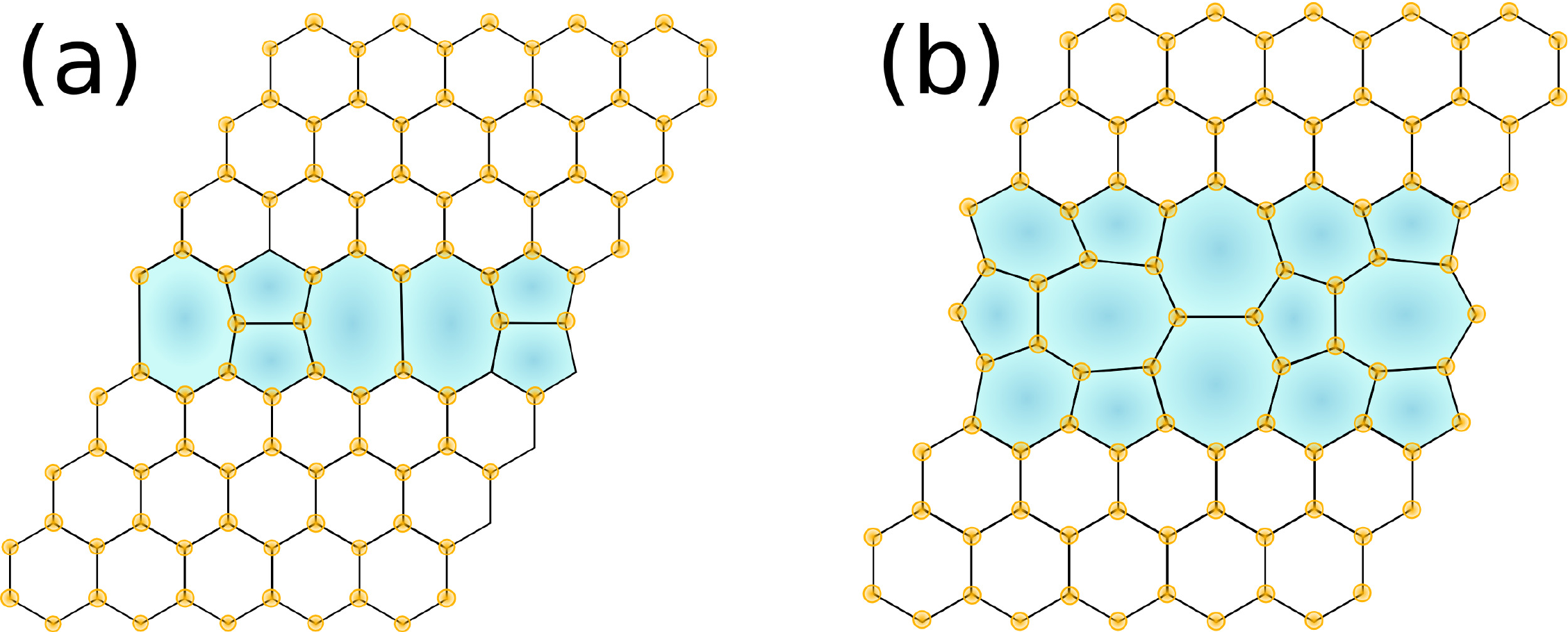}
  \caption{Scheme of two grain boundaries proposed in the context of
    ab-initio works both on graphene and on boron nitride. (a) the
    7557 defect line\cite{Ansari_PCCP:2014} and (b) the t7t5 defect
    line.\cite{Botello-Mendez_Nanoscale:2011} In these schemes we
    highlight in blue the region of the grain boundary.}
  \label{fig:Lit3PdfctLs}
\end{figure}

\begin{figure}[hpt!]
  \includegraphics[width=0.98\columnwidth]{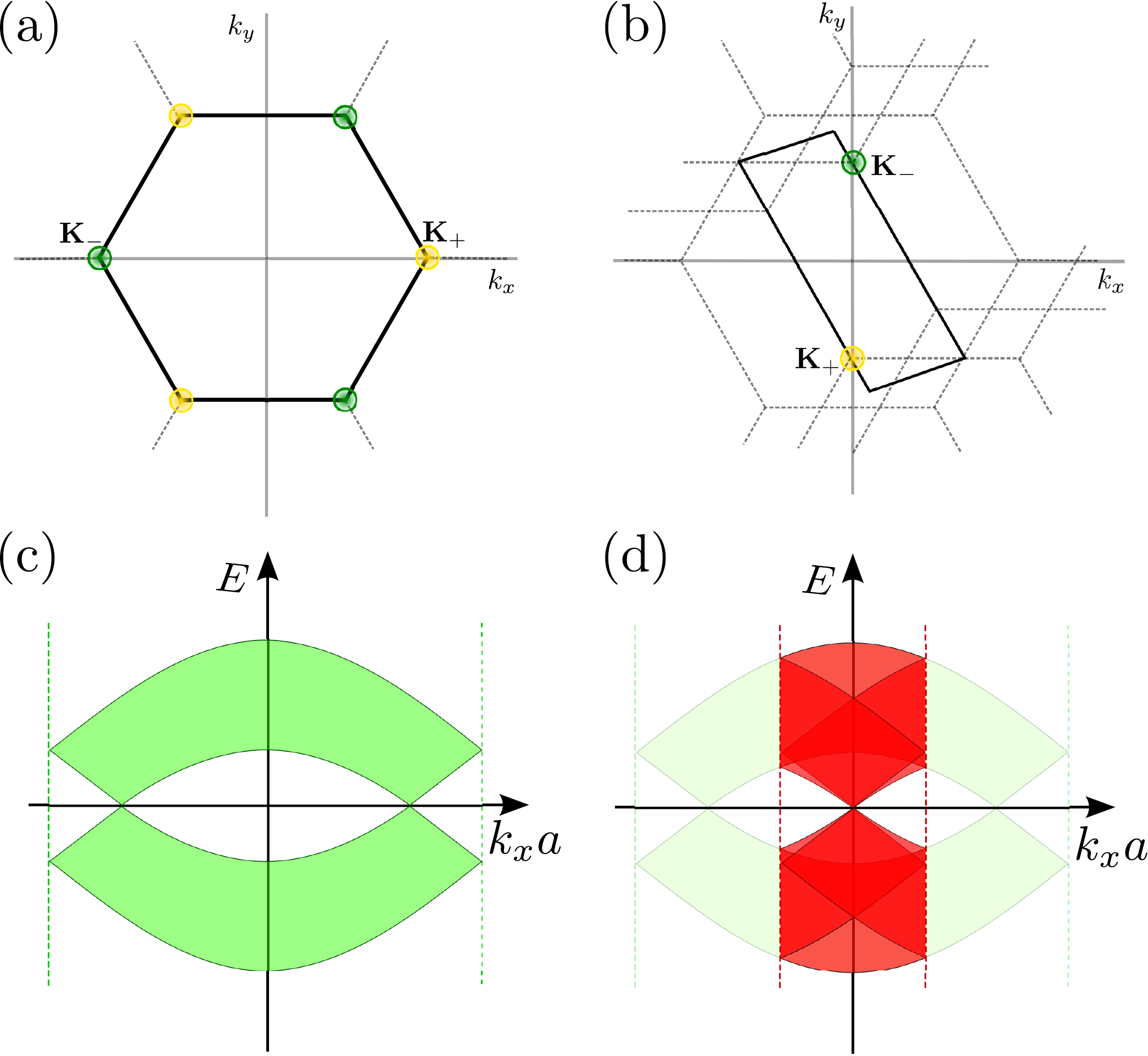}
  \caption{ (a) FBZ of pristine graphene (whose direct lattice vectors
    can be chosen to be $\mathbf{u}_{1} = a (1, 0)$ and
    $\mathbf{u}_{2} = a (-1,\sqrt{3}) / 2$). (b) FBZ of graphene whose
    unit cell has thrice the size of that of pristine graphene in the
    zigzag direction (direction of $\mathbf{u}_{1}$). In this case,
    the direct lattice vectors can be chosen to be $3 \mathbf{u}_{1}$
    and $\mathbf{u}_{2}$ (as done in
    Figs. \ref{fig:DfcLn_7557}-\ref{fig:DfcLn_t7t5}). (c) Spectrum of
    pristine graphene projected along the $k_{x}$ direction (parallel
    to the GB). (d) Comparison between the $k_{x}$-projected spectrum
    of (unfolded) pristine graphene (light green) and the same
    spectrum after a triple folding (red): in the latter, the two
    valleys (and the $\Gamma$-point) are mapped into $k_{x} a = 0$.}
  \label{fig:FBZs}
\end{figure}

From the diversity of GBs belonging to this class, we have
chosen to investigate some that have been recently suggested in the context of ab-initio works both on graphene and on
boron nitride: the 7557 grain boundary\cite{Ansari_PCCP:2014} and the t7t5 grain boundary\cite{Botello-Mendez_Nanoscale:2011}
(see \Fref{fig:Lit3PdfctLs}).
As we will see ahead, these GBs controllably scatter electrons from one valley to the other and can thus be regarded as a
useful nanostructure for valleytronics circuits, whenever the desire is to destroy valley polarization.

A periodic GB preserves the crystal's translation invariance along the GB direction. Therefore, by Fourier transforming the
system's tight-binding Hamiltonian along this direction, we can cast the problem of electronic transport in a 2D crystal
with a GB at its center, as a problem of electronic transport on a quasi-$1$D crystal with a localized defect at its center.

In order to work out this problem we will proceed as in Ref. \onlinecite{RodPS13}. From pristine graphene's Hamiltonian we
start by writing the tight-binding equations away from the grain boundary (see Figs. \ref{fig:DfcLn_7557} and
\ref{fig:DfcLn_t7t5} for notation clarification)
\begin{subequations} \label{eq:BulkTBeqs}
  \begin{eqnarray}
    -\frac{\epsilon}{t}\mathbf{A}(n) &=& W_{A}^{\dagger}\mathbf{B}(n-1)+\mathbf{B}(n),\\
    -\frac{\epsilon}{t}\mathbf{B}(n) &=& \mathbf{A}(n)+W_{A}\mathbf{A}(n+1),
  \end{eqnarray}
\end{subequations}
where we have used the notation $\mathbf{Z}(n)=[Z_{1}(n),Z_{2}(n),Z_{3}(n)]^{T}$ for $Z=A,B$ (sub-lattice identifier). Note that
the notation is hiding the dependency of the $A_{i}$ and $B_{i}$ on $k_{x}$, the momentum along the grain boundary direction. In
the above expressions, $\epsilon$ and $t$ stand respectively for the energy and pristine graphene's hopping parameter, while
$n$ gives the number of direct lattice vector $\mathbf{u}_{2} = (-1,\sqrt{3}) a/2$ translations away from the defect (see Figs.
\ref{fig:DfcLn_7557} and \ref{fig:DfcLn_t7t5}). The matrix $W_{A}$ reads
\begin{eqnarray}
  W_{A} &=& \left[\begin{array}{ccc}
      1 & 1 & 0 \\
      0 & 1 & 1 \\
      e^{3 i k_{x}a} & 0 & 1
    \end{array}\right] .
\end{eqnarray}

We can write the above tight-binding equations in the form
\begin{subequations} \label{eq:DfcTBeqsBulk}
  \begin{eqnarray}
    \left[\begin{array}{c}
        \mathbf{B}(n)\\
        \mathbf{A}(n)
      \end{array}\right] &=& \mathbb{Q}_{1} \left[\begin{array}{c}
        \mathbf{A}(n)\\
        \mathbf{B}(n-1)
      \end{array}\right],\label{eq:Bulk1}\\
    \left[\begin{array}{c}
        \mathbf{A}(n)\\
        \mathbf{B}(n-1)
      \end{array}\right] &=& \mathbb{Q}_{2} \left[\begin{array}{c}
        \mathbf{B}(n-1)\\
        \mathbf{A}(n-1)
      \end{array}\right] , \label{eq:Bulk2}
  \end{eqnarray}
\end{subequations}
where the matrices $\mathbb{Q}_{1}$ and $\mathbb{Q}_{2}$ read
\begin{subequations} \label{eq:Q1Q2}
  \begin{eqnarray}
    \mathbb{Q}_{1} &=& -\left[\begin{array}{cc}
        \frac{\epsilon}{t}\mathbb{I}_{3} & W_{A}^{\dagger}\\
        -\mathbb{I}_{3} & 0
      \end{array}\right] , \\
    \mathbb{Q}_{2} &=& -\left[\begin{array}{cc}
        \frac{\epsilon}{t}\big(W_{A}\big)^{-1} & \big(W_{A}\big)^{-1}\\
        -\mathbb{I}_{3} & 0
      \end{array}\right] ,
  \end{eqnarray}
\end{subequations}
with $\mathbb{I}_{3}$ standing for the $3 \times 3$ unit matrix.

Eqs. (\ref{eq:DfcTBeqsBulk}) can be written in the form of a transfer
matrix equation\cite{RodPL12,RodPS13} relating amplitudes at the atoms
of the unit cell located at $(n-1) \mathbf{u}_{2}$ with the amplitudes
at the atoms of the unit cell located at $n \mathbf{u}_{2}$. Such an
equation reads
\begin{eqnarray}
  \mathbf{L}(n) &=& \mathbb{T}(\epsilon,k_{x}) \mathbf{L}(n-1) , \label{eq:transfMatEq}
\end{eqnarray}
with $\mathbf{L}(n) =
[A_{1}(n),B_{1}(n),A_{2}(n),B_{2}(n),A_{3}(n),B_{3}(n)]^{T}$, and the
transfer matrix, $\mathbb{T}(\epsilon,k_{x})$, given by
\begin{eqnarray}
  \mathbb{T}(\epsilon,k_{x}) &=& R . \mathbb{Q}_{1} . \mathbb{Q}_{2} . R^{T} . \label{eq:TransferMatrixCoup}
\end{eqnarray}
In the above equation, matrix $R$ is simply used to change from the
basis $\{ B_{1},B_{2},B_{3},A_{1},A_{2} ,A_{3} \}$ to the basis $\{
A_{1},B_{1},A_{2},B_{2},A_{3},B_{3} \}$. It is written in
Eq. (\ref{eq:Rmatrix}).

Following the method used for the cases of the 585 and
5757 defect lines,\cite{RodPL12,RodPS13} we can find a basis
where the transfer matrix becomes block diagonal with three $2 \times
2$ matrices on its diagonal. In this basis the three modes of the
problem are uncoupled. Moreover, around $k_{x} = 0$ two of these
modes are low-energy (corresponding to each of the two Dirac cones),
while the other is a high-energy mode.

We can understand this fact from \Fref{fig:FBZs}(b) where one
represents the First Brillouin zone (FBZ) originating from a honeycomb
lattice whose direct vectors are chosen to be $3 \mathbf{u}_{1}$ and
$\mathbf{u}_{2}$. In such a FBZ, the two Dirac points are located at
the same value of $k_{x}$, i. e., at $k_{x} = 0$. It is thus natural
that when setting $k_{x} = 0$ in the transfer matrix given by
Eq. (\ref{eq:TransferMatrixCoup}), one obtains a transfer matrix that
describes simultaneously low-energy electrons at each of the two
valleys (together with an additional high-energy mode associated with
the $\Gamma$-point region of the spectrum of pristine graphene).

The symbol $\Lambda(k_{x})$ stands for the matrix mediating the change to
the basis uncoupling the modes of the transfer matrix
\begin{eqnarray}
  \widetilde{\mathbf{L}}(n) &=& \Lambda(k_{x}) \mathbf{L}(n) .
\end{eqnarray}
We will denote the states in this new basis as
\begin{eqnarray}
  \widetilde{\mathbf{L}} &=& [ A_{h},B_{h},A_{l-},B_{l-},A_{l+},B_{l+} ] ,
\end{eqnarray}
with $h$ identifying the mode with high-energy when
$k_{x}\approx0$, while $l+$ and $l-$ stand for the two modes
with low-energy when $k_{x}\approx0$, one associated with the
$\mathbf{K}_{-}$ valley and the other with the $\mathbf{K}_{+}$
valley. The matrix $\Lambda(k_{x})$ is explicitly written in
Eq. (\ref{eq:LambdaMat}).

As previously stated, in this basis the transfer matrix,
$\widetilde{\mathbb{T}}(\epsilon,k_{x})$, is block diagonal and reads
\begin{eqnarray}
  \widetilde{\mathbb{T}}(\epsilon,k_{x}) &=& \left [ \begin{array}{ccc} \mathbb{T}_{h}(\epsilon,k_{x})
      & 0 & 0 \\ 0 & \mathbb{T}_{l-}(\epsilon,k_{x}) & 0 \\ 0 & 0 & \mathbb{T}_{l+}(\epsilon,k_{x})
    \end{array} \right ] , \label{eq:TransMat}
\end{eqnarray}
where the three transfer matrices associated with each of the
uncoupled modes are written in Eqs. (\ref{eq:TMdiags}).

\subsection{The transmittance across the 7557 and the t7t5 grain boundaries}

\label{sec:Transm}

In a similar manner, the tight-binding Hamiltonian describing the electronic structure close to the grain boundary
can be used to write the tight-binding equations for the defect. With these we can compute a boundary condition
relating amplitudes on either side of the defect
\begin{eqnarray}
  \mathbf{L}(1) &=& \mathbb{M}.\mathbf{L}(-1), \label{eq:TBbc}
\end{eqnarray}
where the boundary condition matrix, $\mathbb{M}$, is a $6\times6$ matrix that depends both on the energy $\epsilon$,
$x$-momentum $k_{x}$ and the electron hoppings characteristic of the grain boundary. In Appendix
\ref{app:abinitioDfcLns_bc} we compute these matrices for the two grain boundaries we are investigating: the 7557 and
the t7t5 grain boundaries (see Figs. \ref{fig:DfcLn_7557} and \ref{fig:DfcLn_t7t5}).

Note that by expressing this boundary condition matrix in the basis that uncouples the modes of the transfer matrix
$\mathbb{T}(\epsilon,k_{x})$
\begin{eqnarray}
  \widetilde{\mathbb{M}} &=& \Lambda(k_{x}a) . \mathbb{M} . [\Lambda(k_{x}a)]^{-1} \, , \label{eq:TBbc_propBasis}
\end{eqnarray}
we can conclude that in general, it mixes all the three modes of matrix $\mathbb{T}(\epsilon,k_{x})$.

Given this, we now have all the ingredients needed to compute the coefficients involved in the electronic scattering
by such defects. An incoming electronic wave from $n=-\infty$, will be scattered by the defect at $n=0$ producing a
reflected and a transmitted component. The wave-function on each side of the defect will then be given by
\begin{subequations} \label{eq:WFs}
  \begin{eqnarray}
    \widetilde{\mathbf{L}}(n<0) &=& \lambda_{i>}^{n+1} \, \mathbf{\Psi}_{i}^{>} + \sum_{j=1}^{r_{<}} \, \rho_{ij} \, \lambda_{j<}^{n+1} \,
    \mathbf{\Psi}_{j}^{<} \,, \label{eq:WFlower} \\
    \widetilde{\mathbf{L}}(n>0) &=& \sum_{j=1}^{r_{>}} \, \tau_{ij} \, \lambda_{j>}^{n-1} \, \mathbf{\Psi}_{j}^{>} \,, \label{eq:WFright}
  \end{eqnarray}
 \end{subequations}
where $\rho_{i j}$ and $\tau_{i j}$ are, respectively, the reflection and transmission scattering amplitudes from an incoming
(from $n=-\infty$) state, $\mathbf{\Psi}_{i}^{>}$, into reflected, $\mathbf{\Psi}_{j}^{<}$, and transmitted, $\mathbf{\Psi}_{j}^{>}$,
outgoing states. Finally, by imposing the corresponding boundary condition [see Eqs. (\ref{eq:TBbc}) and (\ref{eq:TBbc_propBasis})],
we can compute the coefficients $\rho_{i j}$ and $\tau_{i j}$ for a given energy and a given longitudinal momentum.

For both the t7t5 and the 7557 grain boundary we have set the hopping parameters in the region of the grain boundary
by estimating the corresponding carbon-carbon distances originating from the ab-initio results of Refs.
\onlinecite{Botello-Mendez_Nanoscale:2011} and \onlinecite{Ansari_PCCP:2014}, and then using the
parametrization\cite{Tang_PRB:1996}
\begin{eqnarray}
\tau(r_{ij})=\big(\frac{r_{ij}}{a_{0}}\big)^{-\alpha_{2}}\exp[-\alpha_{3}\times(r_{ij}^{\alpha_{4}}-a_{0}^{\alpha_{4}})],\label{eq:hoppParam}
\end{eqnarray}
where $r_{ij}$ stands for the distance between the carbons labeled by $i$ and $j$ (given in units of angstroms), the
adimensional parameters $\alpha_{2}=1.2785$, $\alpha_{3}=0.1383$, $\alpha_{4}=3.4490$, while $a_{0}$ is the carbon-carbon
distance in pristine graphene (in units of angstroms).

\begin{figure}[hpt!]
  \includegraphics[width=0.98\columnwidth]{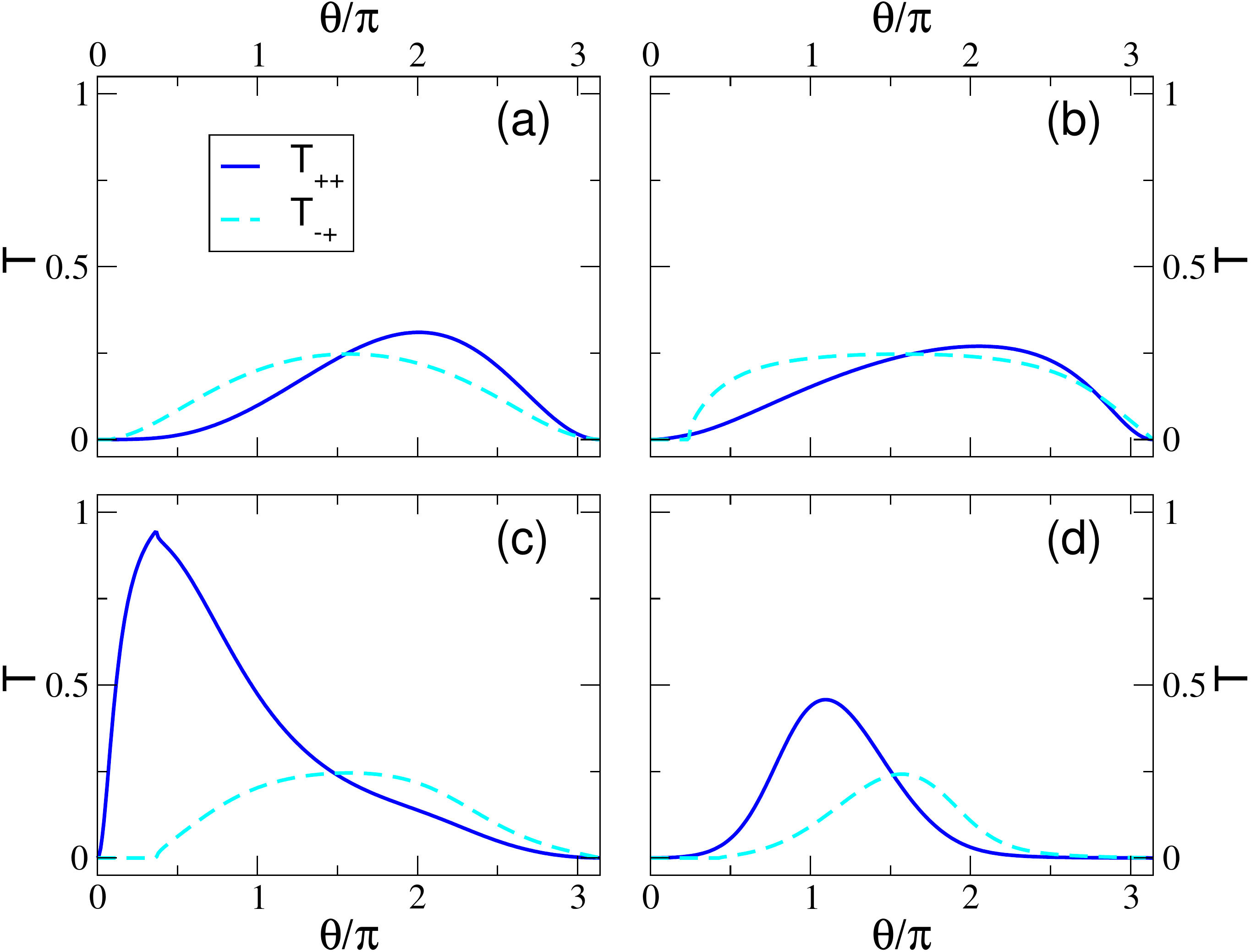}
  \caption{Transmittance in terms of the incidence angle (incoming
    electron chosen to be on the valley $\mathbf{K}_{+}$ Dirac point)
    for the 7557 grain boundary. The panels (a), (b), (c) and (d)
    correspond to scattering processes occurring at energies of,
    respectively, $0.01 t$, $0.1 t$, $0.3 t$ and $0.5 t$. The hopping
    parameters (see Fig. \ref{fig:DfcLn_7557}) were set at:
    $\xi=0.98$, $\gamma=0.94$ and $\beta=0.1$. The dark blue curve
    stands for the transmittance preserving the valley degree of
    freedom (electron from the $\mathbf{K}_{+}$ below the GB
    scattering scatters to the same valley above the GB, i. e. $T_{++}
    = \vert \tau_{++} \vert^{2}$). The dashed light blue curve stands
    for the intervalley transmittance (electron at the
    $\mathbf{K}_{+}$ valley below the GB scattering into the valley
    $\mathbf{K}_{-}$ above the GB, i. e. $T_{-+} = (v_{-}/v_{+}) \,
    \vert \tau_{-+} \vert^{2}$; $v_{\pm}$ stands for the velocity of
    the mode $l\pm$).}
  \label{fig:Trans_7557}
\end{figure}

In Fig. \ref{fig:Trans_7557} we present the transmission probability for the 7557 grain boundary (see scheme of Fig.
\ref{fig:DfcLn_7557} and Appendix \ref{app:7557}) of an incoming electron of the $\mathbf{K}_{+}$ valley. The several
transmittance curves of this figure correspond to different energies and were drawn using the following hopping
parameters at the defect: $\xi=0.98$, $\gamma=0.94$ and $\beta=0.1$.
One can see on the several panels of this figure that the intervalley scattering is comparable to the {\it valley-preserving}
scattering. Both of them strongly depend on the energy and incidence angle, mainly due to the dependence on energy and $k_{x}$
of the boundary condition matrix (see its computation in Appendix \ref{app:7557}).

\begin{figure}[hpt!]
  \includegraphics[width=0.98\columnwidth]{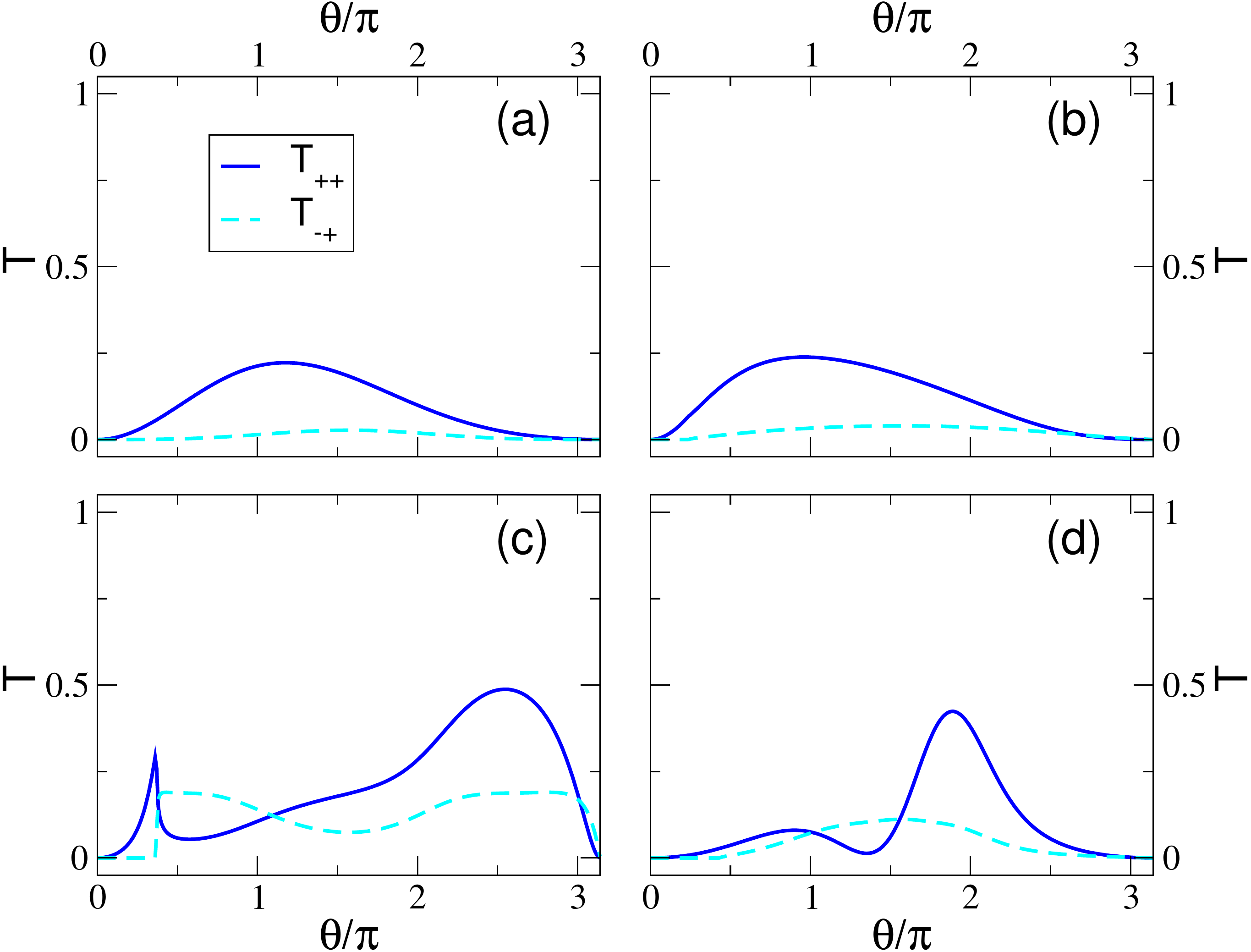}
  \caption{Transmittance in terms of the incidence angle (incoming
    electron chosen to be on the valley $\mathbf{K}_{+}$ Dirac point)
    for the t7t5 grain boundary. The panels (a), (b), (c) and (d)
    correspond to scattering processes occurring at energies of,
    respectively, $0.01 t$, $0.1 t$, $0.3 t$ and $0.5 t$. The hopping
    parameters (see Fig. \ref{fig:DfcLn_t7t5}) were set at:
    $\xi_{1}=1.06$, $\xi_{2}=0.95$, $\xi_{3}=0.83$, $\xi_{4}=0.80$,
    $\xi_{5}=1.30$, $\xi_{6}=1.05$, $\gamma_{1}=1.23$,
    $\gamma_{2}=1.20$, $\gamma_{3}=1.18$ and $\gamma_{1}=1.36$. The
    dark blue curve stands for the transmittance preserving the valley
    degree of freedom (electron from the $\mathbf{K}_{+}$ below the GB
    scattering scatters to the same valley above the GB, i. e. $T_{++}
    = \vert \tau_{++} \vert^{2}$). The dashed light blue curve stands
    for the intervalley transmittance (electron at the
    $\mathbf{K}_{+}$ valley below the GB scattering into the valley
    $\mathbf{K}_{-}$ above the GB, i. e. $T_{-+} = (v_{-}/v_{+}) \,
    \vert \tau_{-+} \vert^{2}$; $v_{\pm}$ stands for the velocity of
    the mode $l\pm$).}
  \label{fig:Trans_t7t5}
\end{figure}

Similar plots are presented in Fig. \ref{fig:Trans_t7t5} for the t7t5 grain boundary (see scheme of Fig. \ref{fig:DfcLn_t7t5}
and Appendix \ref{app:t7t5}). These were obtained with the following hopping parameters: $\xi_{1}=1.06$, $\xi_{2}=0.95$,
$\xi_{3}=0.83$, $\xi_{4}=0.80$, $\xi_{5}=1.30$, $\xi_{6}=1.05$, $\gamma_{1}=1.23$, $\gamma_{2}=1.20$, $\gamma_{3}=1.18$ and
$\gamma_{1}=1.36$.
In general, this choice of hopping parameters gives rise to a lower intervalley scattering at low energies than
what is obtained for the 7557 grain boundary.

We finalize by noting that the above transmittance curves are strongly dependent on the choice of the hopping parameters.
In particular, at low energies these are the only parameters determining the boundary condition matrix and therefore
controlling the system's transparency to incident electrons. Moreover, the GB's scattering profile can be
strongly enhanced or suppressed by small changes of the GB's hopping parameters. Therefore, we may expect that intervalley
scattering at the GB is deeply sensitive to modifications of the lattice's geometry (namely strain) in the vicinity of the 
GB.

\begin{acknowledgments}
NMRP acknowledges support from EC under Graphene Flagship (Contract No.
CNECT-ICT-604391) and the hospitality of the Instituto de F\'{\i}sica of the UFRJ, where this work
was completed. JNBR acknowledges Singapore National Research Foundation for its support through the
Fellowship Program NRF-NRFF2012-01. CJP and ALCP acknowledge São Paulo Research Foundation (FAPESP), grant 2012/19060-0.
Part of the numerical simulations were performed at the computational facilities
from CENAPAD-SP, at Campinas State University.
\end{acknowledgments}

%
%
\appendix

\section{The bulk tight-binding equations}

\label{app:TBeqs}

The matrix $R$ changing from the basis $\{B_{1}(n),B_{2}(n),B_{3}(n),A_{1}(n),A_{2}(n),A_{3}(n) \}$ into the basis
$\{ A_{1}(n),B_{1}(n),A_{2}(n),B_{2}(n),A_{3}(n),B_{3}(n) \}$ reads
\begin{eqnarray}
  R &=& \left[\begin{array}{cccccc}
      0 & 0 & 0 & 1 & 0 & 0 \\
      1 & 0 & 0 & 0 & 0 & 0 \\
      0 & 0 & 0 & 0 & 1 & 0 \\
      0 & 1 & 0 & 0 & 0 & 0 \\
      0 & 0 & 0 & 0 & 0 & 1 \\
      0 & 0 & 1 & 0 & 0 & 0
    \end{array}\right] . \label{eq:Rmatrix}
\end{eqnarray}

The matrix mediating the basis change that uncouples the modes of the
transfer matrix, $\Lambda(k_{x})$, reads
\begin{widetext}
  \begin{eqnarray}
    \Lambda\Big(\frac{\phi}{a}\Big) &=& \frac{1}{\sqrt{3}} \left [ \begin{array}{cccccc} 1 & 0 & - \frac{e^{-i (\phi - 2 \pi/3)}
          i \sqrt{3}}{1 + e^{i \pi/3}} & 0 & \frac{e^{-i 2 (\phi - 2 \pi/3)} i \sqrt{3}}{1 + e^{-i \pi/3}} & 0 \\
        0 & 1 & 0 & - \frac{e^{-i (\phi - 2 \pi/3)} i \sqrt{3}}{1 + e^{i \pi/3}} & 0 & \frac{e^{-i 2 (\phi - 2 \pi/3)}
          i \sqrt{3}}{1 + e^{-i \pi/3}} \\
        1 & 0 & -e^{-i (\phi - \pi/3)} & 0 & -e^{-i (2 \phi + \pi/3)} & 0 \\
        0 & 1 & 0 & -e^{-i (\phi - \pi/3)} & 0 & -e^{-i (2 \phi + \pi/3)} \\
        1 & 0 & -e^{-i (\phi + \pi/3)} & 0 & -e^{-i (2 \phi - \pi/3)} & 0 \\
        0 & 1 & 0 & -e^{-i (\phi + \pi/3)} & 0 & -e^{-i (2 \phi - \pi/3)}
      \end{array} \right ] , \label{eq:LambdaMat}
  \end{eqnarray}
\end{widetext}
where $\phi = k_{x} a$.

As just said, in this basis the transfer matrix,
Eq. (\ref{eq:TransferMatrixCoup}), becomes block diagonal with three
$2 \times 2$ matrices in its diagonal. The three pairs of modes, $h$,
$l+$ and $l-$, decouple and propagate independently. If we put
ourselves around the Dirac point $\mathbf{K}_{+} = (0,-1) \nu 4 \pi/(3
\sqrt{3}a)$, the upper matrix corresponds to the high-energy mode, the
middle one corresponds to the Dirac cone identified by $\nu=-1$, while
the lower matrix stands for the cone identified by $\nu=+1$. For a
general energy and momentum these three matrices read
\begin{subequations} \label{eq:TMdiags}
  \begin{eqnarray}
    \mathbb{T}_{h}(\epsilon,\phi) &=& \frac{1}{1 + e^{i \phi}} \left [ \begin{array}{cc} -1 & - \epsilon
        \\ \epsilon & \epsilon^{2} -2 - 2 \cos\phi \end{array} \right ] \,, \label{eq:TMdiags1} \\
    \mathbb{T}_{l-}(\epsilon,\phi) &=& f(\phi) \left [ \begin{array}{cc} - 1 & - \epsilon \\ \epsilon
        & \epsilon^{2} + \frac{e^{-i (\phi - \frac{\pi}{3})}- 1}{f(\phi)} \end{array} \right ] \,, \label{eq:TMdiags2} \\
    \mathbb{T}_{l+}(\epsilon,\phi) &=& g(\phi) \left [ \begin{array}{cc} -1 & -\epsilon \\ \epsilon &
         \epsilon^{2} + \frac{e^{-i (\phi + \frac{\pi}{3})}-1}{g(\phi)} \end{array} \right ] \,, \label{eq:TMdiags3}
  \end{eqnarray}
\end{subequations}
where we have again used $\phi=k_{x} a$ and have defined $f(\phi)$ and $g(\phi)$ as
\begin{subequations}
  \begin{eqnarray}
    f(\phi) &=& \frac{e^{i \pi / 3} - e^{-i \phi}}{1 - 2 \cos\phi} \,, \\
    g(\phi) &=& \frac{e^{-i \pi / 3} - e^{-i \phi}}{1 - 2 \cos\phi} \,.
  \end{eqnarray}
\end{subequations}

\section{The boundary condition of the 7557 and t7t5 grain boundaries}

\label{app:abinitioDfcLns_bc}

In this appendix we will briefly compute the boundary condition matrix associated with the two grain boundaries
investigated in Section \ref{sec:TransfMatrixFormalism}.

\subsection{The boundary condition of the 7557 grain boundary}

\label{app:7557}

\begin{figure}
  \includegraphics[width=0.98\columnwidth]{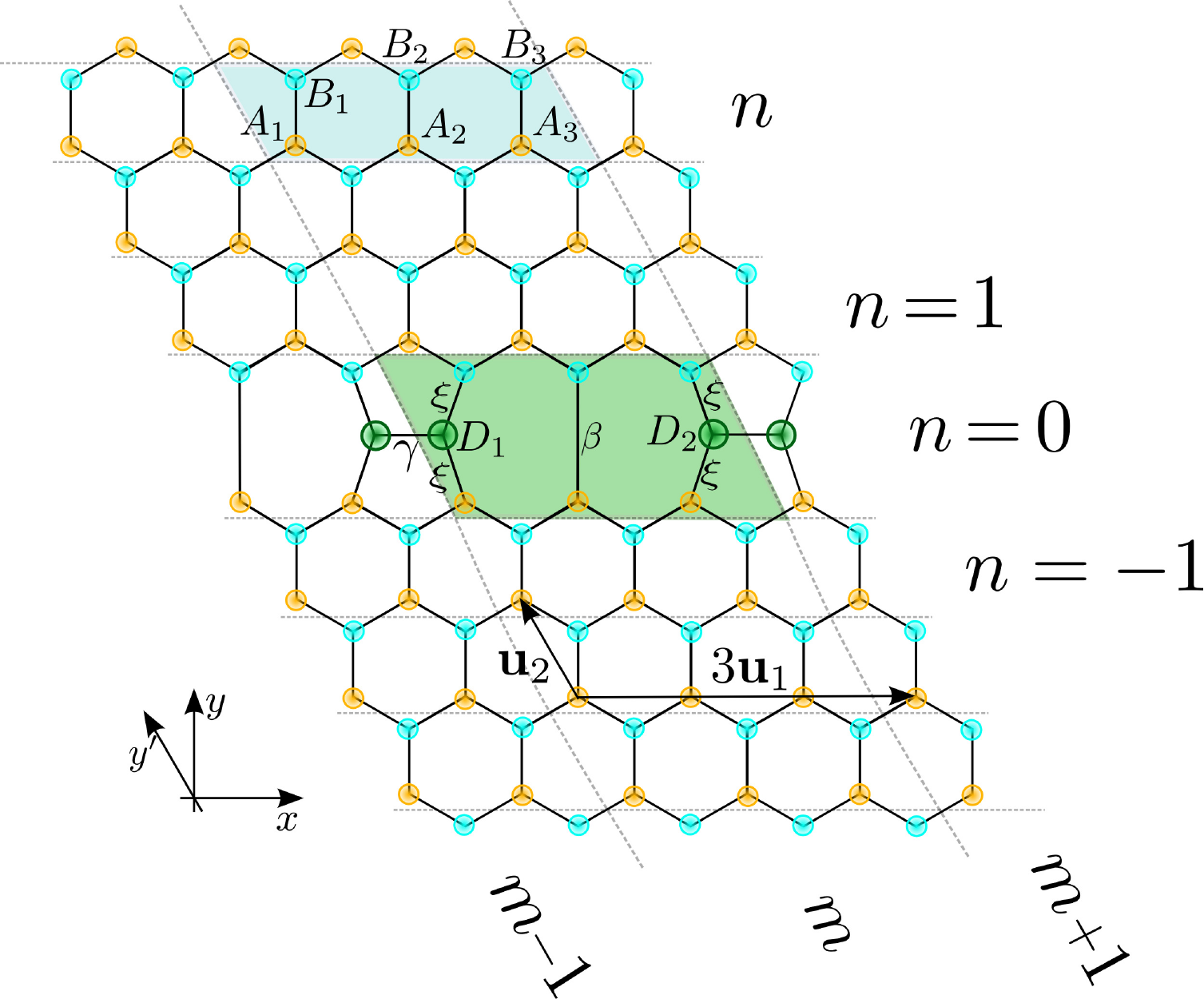}
  \caption{Crystalline structure of the 7557 grain
    boundary.\cite{Ansari_PCCP:2014} The region of the defect line is
    highlighted in blue.}
  \label{fig:DfcLn_7557}
\end{figure}

Let us start by computing the boundary condition matrix relating the wave-function amplitudes on either side
of the 7557 grain boundary (see Fig. \ref{fig:DfcLn_7557} for a scheme of its crystalline structure).
The tight-binding equations at the defect region read
\begin{subequations} \label{eq:DfcTBeqs-7557}
  \begin{eqnarray}
    -\frac{\epsilon}{t}\mathbf{B}(0) &=& \mathbf{A}(1) + \mathcal{X}^{T} \mathbf{D} + \mathcal{B} \mathbf{A}(0),\\
    -\frac{\epsilon}{t}\mathbf{D} &=& \mathcal{X} \big( \mathbf{A}(0)+\mathbf{B}(0) \big) + \mathcal{G} \mathbf{D},\\
    -\frac{\epsilon}{t}\mathbf{A}(0) &=& \mathcal{X}^{T} \mathbf{D} + W_{A}^{\dagger} \mathbf{B}(-1) + \mathcal{B} \mathbf{B}(0),
  \end{eqnarray}
\end{subequations}
where we use the notation $\mathbf{Z}(n)=[Z_{1}(n),Z_{2}(n),Z_{3}(n)]^{T}$ with $Z=A,B,D$. The matrices $\mathcal{X}$,
$\mathcal{B}$ and $\mathcal{G}$ read
\begin{subequations}
  \begin{eqnarray}
    \mathcal{X} &=& \left[\begin{array}{ccc}
        \xi & 0 & 0 \\
        0 & 0 & \xi \\
        0 & 0 & 0
      \end{array}\right] \,, \label{eq:Xx} \\
    \mathcal{B} &=& \left[\begin{array}{cccc}
        0 & 0 & 0 \\
        0 & \beta & 0 \\
        0 & 0 & 0 \\
      \end{array}\right] \,, \label{eq:Bb} \\
    \mathcal{G} &=& \left[\begin{array}{ccc}
        0 & \gamma e^{-i 3 \phi} & 0 \\
        \gamma  e^{i 3 \phi} & 0 & 0 \\
        0 & 0 & 0 \\
      \end{array}\right] \,, \label{eq:Gg}
  \end{eqnarray}
\end{subequations}
where $\phi = k_{x} a$.

The above equations give the boundary condition between either side of the grain boundary which reads
\begin{eqnarray}
  \left[\begin{array}{c}
      \mathbf{B}(1)\\
      \mathbf{A}(1)
    \end{array}\right] &=&
  \mathbb{M}_{1} . \mathbb{M}_{2} \left[\begin{array}{c}
      \mathbf{B}(-1)\\
      \mathbf{A}(-1)
    \end{array}\right] \,, \label{eq:755t-bcmatrix}
\end{eqnarray}
where the $\mathbb{N}_{i}$ are $6 \times 6$ matrices that read
\begin{subequations} \label{eq:NmatricesDfcLn-7557}
  \begin{eqnarray}
    \mathbb{M}_{1} &=& \left[\begin{array}{cc}
        \frac{\epsilon}{t} \mathcal{P} - W_{A}^{\dagger} & \frac{\epsilon}{t} \mathcal{Q} \\
        - \mathcal{P} & - \mathcal{Q}
      \end{array}\right] \,, \\
    \mathbb{M}_{2} &=& \left[\begin{array}{cc}
        \mathcal{Q}^{-1} \big( \frac{\epsilon}{t} \mathcal{P} W_{A}^{-1} - W_{A}^{\dagger} \big) & \mathcal{Q}^{-1} \mathcal{P} W_{A}^{-1} \\
        - \frac{\epsilon}{t} W_{A}^{-1} & -W_{A}^{-1}
      \end{array}\right] \,.
  \end{eqnarray}
\end{subequations}
In Eqs. (\ref{eq:NmatricesDfcLn-7557}) we have used the following definitions for the matrices $\mathcal{P}$ and $\mathcal{Q}$,
\begin{subequations} \label{eq:FuncsNmatricesDfcLn1}
  \begin{eqnarray}
    \mathcal{P} &=& \frac{\epsilon}{t} \mathbb{I}_{3} - \mathcal{X}^{T} \mathcal{R} \,, \\
    \mathcal{Q} &=& \mathcal{B} + \mathcal{X}^{T} \mathcal{R} \,,
  \end{eqnarray}
\end{subequations}
where $\mathbb{I}_{3}$ stands for the $3 \times 3$ identity matrix, while the matrix $\mathcal{R}$ reads
\begin{eqnarray}
  \mathcal{R} &=& - \frac{1}{\xi ^2} \left[ \begin{array}{ccc}
      \epsilon & 0 & e^{-3 i \phi } \gamma \\
      0 & -\frac{\xi^{2}}{\beta} & 0 \\
      e^{3 i \phi } \gamma & 0 & \epsilon \\
    \end{array} \right] \,.
\end{eqnarray}
Note that the above matrices depend on the reduced energy, $\epsilon/t$, the longitudinal momentum, $k_{x}$, and the
hopping parameters at the defect, $\xi$, $\gamma$ and $\beta$. Similarly, the boundary condition connecting the two
sides of the defect, i. e., $\mathbf{L}(1) = \mathbb{M}_{7557} . \mathbf{L}(-1)$, reads
\begin{eqnarray}
  \mathbb{M}_{7557} &=& R . \mathbb{M}_{1} . \mathbb{M}_{2} . R^{T} , \label{eq:TBDCMat-7557}
\end{eqnarray}
and in general depends on $\xi$, $\gamma$, $\beta$, $\epsilon/t$ and $k_{x}$.

\subsection{The boundary condition of the t7t5 grain boundary}

\label{app:t7t5}

\begin{figure}
  \includegraphics[width=0.98\columnwidth]{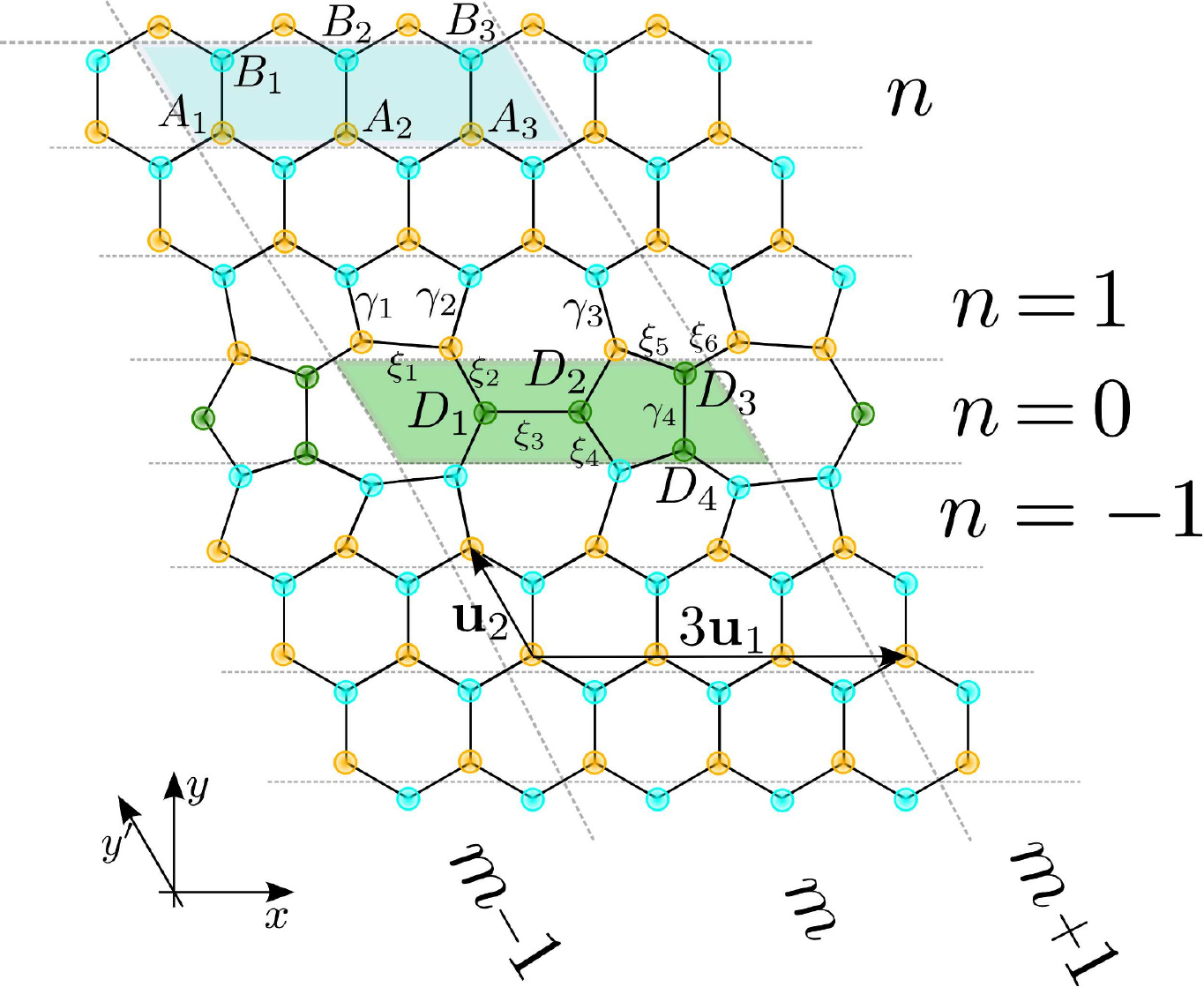}
  \caption{Crystalline structure of the t7t5 defect
    line.\cite{Botello-Mendez_Nanoscale:2011} The region of the defect
    line is highlighted in blue.}
  \label{fig:DfcLn_t7t5}
\end{figure}

In Fig. \ref{fig:DfcLn_t7t5} we can see the scheme of the crystalline structure of a t7t5 grain boundary.
In what follows we will compute the boundary condition matrix relating the wave-function amplitudes on either of its sides.
For such a grain boundary, the tight-binding equations in the defect region read
\begin{subequations} \label{eq:DfcTBeqs}
  \begin{eqnarray}
    -\frac{\epsilon}{t}\mathbf{B}(1) &=& G_{+}\mathbf{A}(1)+W_{A}\mathbf{A}(2),\\
    -\frac{\epsilon}{t}\mathbf{A}(1) &=& G_{+}\mathbf{B}(1)+X_{1}\mathbf{A}(1)+X_{2}\mathbf{D},\\
    -\frac{\epsilon}{t}\mathbf{D} &=& X_{2}^{\dagger}\mathbf{A}(1)+\mathcal{S}\mathbf{D}+X_{3}\mathbf{B}(-1),\\
    -\frac{\epsilon}{t}\mathbf{B}(-1) &=& G_{-}\mathbf{A}(-1)+X_{3}^{\textrm{T}}\mathbf{D}+X_{4}\mathbf{B}(-1),\\
    -\frac{\epsilon}{t}\mathbf{A}(-1) &=& G_{-}\mathbf{B}(-1)+W_{A}^{\dagger}\mathbf{B}(-2),
  \end{eqnarray}
\end{subequations}
where, once more we use the notation $\mathbf{Z}(n)=[Z_{1}(n),Z_{2}(n),Z_{3}(n)]^{T}$, now for $Z=A,B,D$. The $G_{\pm}$ are diagonal
matrices that can be written as $G_{+} = \textrm{diag}[\gamma_{1},\gamma_{2},\gamma_{3}]$ and $G_{-} = \textrm{diag}[\gamma_{2},\gamma_{3},
\gamma_{1}]$, while the $X_{i}$ matrices (with $i=1,2,3,4$) read
\begin{eqnarray}
  X_{1} &=& \xi_{1} \left[\begin{array}{ccc}
      0 & 1 & 0 \\
      1 & 0 & 0 \\
      0 & 0 & 0
    \end{array}\right] \,, \label{eq:X1} \\
  X_{2} &=& \left[\begin{array}{cccc}
      0 & 0 & \xi_{6} e^{-i 3 \phi} & 0 \\
      \xi_{2} & 0 & 0 & 0 \\
      0 & \xi_{4} & \xi_{5} & 0 \\
    \end{array}\right] \,, \label{eq:X2} \\
  X_{3} &=& \left[\begin{array}{ccc}
      \xi_{2} & 0 & 0 \\
      0 & \xi_{4} & 0 \\
      0 & 0 & 0 \\
      0 & \xi_{5} & \xi_{6} \\
    \end{array}\right] \,, \label{eq:X3} \\
  X_{4} &=& \xi_{1} \left[\begin{array}{ccc}
      0 & 0 & e^{-i 3 \phi} \\
      0 & 0 & 0 \\
      e^{i 3 \phi} & 0 & 0
    \end{array}\right] \,, \label{eq:X4}
\end{eqnarray}
where again $\phi = k_{x} a$. Finally, $\mathcal{S}$ reads
\begin{eqnarray}
  \mathcal{S} &=& \left[\begin{array}{cccc}
      0 & \xi_{3} & 0 & 0 \\
      \xi_{3} & 0 & 0 & 0 \\
      0 & 0 & 0 & \xi_{4} \\
      0 & 0 & \xi_{4} & 0 \\
    \end{array}\right] \,. \label{eq:RR}
\end{eqnarray}
We can rewrite the above equations in a more compact form that allows us to write the equation relating the amplitudes at
each side of the grain boundary (i. e., those at $n=2$ with those at $n=-2$) in the following way
\begin{eqnarray}
  \left[\begin{array}{c}
      \mathbf{B}(2)\\
      \mathbf{A}(2)
    \end{array}\right] &=&
  \mathbb{N}_{1} . \mathbb{N}_{2} . \mathbb{N}_{3} . \mathbb{N}_{4} . \mathbb{N}_{5} . \mathbb{N}_{6} \left[\begin{array}{c}
      \mathbf{B}(-2)\\
      \mathbf{A}(-2)
    \end{array}\right] \,, \label{eq:t7t5-bcmatrix}
\end{eqnarray}
where the matrices $\mathbb{N}_{i}$ are now $6 \times 6$ reading
\begin{subequations} \label{eq:NmatricesDfcLn}
  \begin{eqnarray}
    \mathbb{N}_{1} &=& -\left[\begin{array}{cc}
        \frac{\epsilon}{t}\mathbb{I}_{3} & \big(W_{A}\big)^{\dagger}\\
        -\mathbb{I}_{3} & 0
      \end{array}\right] \,, \\
    \mathbb{N}_{2} &=& -\left[\begin{array}{cc}
        \frac{\epsilon}{t}\big(W_{A}\big)^{-1} & \big(W_{A}\big)^{-1} G_{+} \\
        -\mathbb{I}_{3} & 0
      \end{array}\right] \,, \\
    \mathbb{N}_{3} &=& -\left[\begin{array}{cc}
        G_{+}^{-1} F_{1} & G_{+}^{-1} X_{2} P^{-1} X_{3} \\
        -\mathbb{I}_{3} & 0
      \end{array}\right] \,,
  \end{eqnarray}
and
  \begin{eqnarray}
    \mathbb{N}_{4} &=& -\left[\begin{array}{cc}
        Q^{-1} F_{2} & Q^{-1} G_{-}^{-1} \\
        -\mathbb{I}_{3} & 0
      \end{array}\right] \,, \\
    \mathbb{N}_{5} &=& -\left[\begin{array}{cc}
        \frac{\epsilon}{t} G_{-}^{-1} & G_{-}^{-1} \big( W_{A} \big)^{\dagger} \\
        -\mathbb{I}_{3} & 0
      \end{array}\right] \,, \\
    \mathbb{N}_{6} &=& -\left[\begin{array}{cc}
        \frac{\epsilon}{t} W_{A}^{-1} & W_{A}^{-1} \\
        -\mathbb{I}_{3} & 0
      \end{array}\right] \,,
  \end{eqnarray}
\end{subequations}
where we have used the following definitions
\begin{subequations} \label{eq:FuncsNmatricesDfcLn}
\begin{eqnarray}
  F_{1} &=& \frac{\epsilon}{t} \mathbb{I}_{3} + X_{1} + X_{2} P^{-1} X_{2}^{\dagger} \,, \\
  F_{2} &=& \frac{\epsilon}{t} \mathbb{I}_{3} + X_{3}^{\textrm{T}} P^{-1} X_{3} + X_{4} \,, \\
  P &=& - \frac{\epsilon}{t} \mathbb{I}_{3} - \mathcal{S} \,, \\
  Q &=& X_{3}^{\textrm{T}} P^{-1} X_{2}^{\dagger} \,,
\end{eqnarray}
\end{subequations}
The above matrices depend on the reduced energy, $\epsilon/t$, the longitudinal momentum, $k_{x}$, and the hopping
parameters at the defect, $\xi_{i}$ and $\gamma_{j}$ (with $i=1,\ldots,6$ and $j=1,\ldots,4$).

It is now straightforward to write the boundary condition connecting the two sides of the defect $\mathbf{L}(1) =
\mathbb{M}_{t7t5} . \mathbf{L}(-1)$, where the boundary condition matrix, $\mathbb{M}$, is a $6\times6$ matrix given by
\begin{eqnarray}
  \mathbb{M}_{t7t5} &=& R . \mathbb{N}_{1} . \mathbb{N}_{2} . \mathbb{N}_{3} . \mathbb{N}_{4} . \mathbb{N}_{5} . \mathbb{N}_{6}
  . R^{T} , \label{eq:TBDCMat-t7t5}
\end{eqnarray}
where, for the sake of simplicity of notation, we have omitted the dependence of the matrices $\mathbb{M}_{t7t5}$ and
$\mathbb{N}_{i}$ on $\epsilon/t$, $k_{x}$, $\xi_{i}$ and $\gamma_{j}$.

\bibliography{bibliograph}

\end{document}